\documentclass{aa}

\usepackage{graphicx}
\usepackage{txfonts}

\begin{document}

   \title{The planetary nebula IC~5148 and its ionized halo\thanks{Based on observations made with ESO Telescopes at the La Silla Paranal Observatory under program ID 098.D-0332}}

   \author{D. Barr{\'i}a\inst{1}
   \and S. Kimeswenger\inst{1,2}
\and W. Kausch\inst{2}
\and D.~S. Goldman\inst{3}
          }

   \institute{
    Instituto de Astronom{\'i}a, Universidad Cat{\'o}lica del Norte, Av. Angamos 0610, Antofagasta, Chile
\and
Institut f{\"u}r Astro- und Teilchenphysik, Leopold--Franzens Universit{\"a}t Innsbruck, Technikerstr. 25, 6020 Innsbruck, Austria
\and
Astrodon Imaging, 407 Tyrolean Court. Roseville, California, 95661, USA}

   \date{Received 30$^{\rm th}$ July, 2018; accepted 2$^{\rm th}$ October, 2018}


  \abstract
   {Many round or nearly roundish Planetary Nebulae (PNe) show multiple shells and halo structures during their evolutionary stage near the maximum temperature of their central star. Controversial debate is currently ongoing if these structures are recombination halos, as suggested by hydrodynamic modelling efforts, or ionized material. Recently we discovered a halo with even somewhat unusual structures around the sparsely studied PN IC~5148 and present for the first time spectroscopy going out to the halo of such a PN.}
   {We investigate the spatial distribution of material and its ionization state from the center of the nebula up to the very outskirts of the halo. }
   {We obtained long-slit low resolution spectroscopy (FORS2@VLT) of the nebula in two position angles, which we used to investigate the nebular structure and its halo in the optical range from 450 to 880\,nm. In addition we used medium resolution spectra taken with X-SHOOTER@VLT ranging from 320\,nm to 2.4\,$\mu$m to derive atmospheric parameters for the central star. We obtained the distance and position in the Galaxy from various methods combined with GAIA DR2 data. We also applied Cloudy models to the nebula in order to derive physical parameters of the various regions.}
   {We obtained spatially resolved structures and detailed descriptions of the outrunning shock front and a set of unusual halo structures denoted to further shock. The halo structures appears clearly as hot ionized material. Furthermore we derived a reliable photometric value for the central star at a GAIA distance of $D=1.3$\,kpc. Considering the large distance $z=1.0$\,kpc from the galactic plane together to its non-circular motion in the galaxy and, a metallicity only slightly below that of typical disk PNe, most likely IC 5148 originates from a thick disk population star.}
   {}


   \maketitle
%

\section{Introduction}

Although the interacting wind model for the formation of planetary
nebulae (PNe) has been accepted for a long time \citep{Kwok78,Kwok82,Kwok02} and morphological sequences were drawn in detail \citep{Balick87}, the exact mechanism is not yet clearly understood up to now. However, significant progress have been made in the past years. Several numerical simulations starting with
\citet{Ok85} followed by \citet{MeFr94,MeFr95} and by the group around Schönberner \citep[e.g.][]{Sch98a,Sch98b,Sch00,Sch04,Sch14}, have been carried out. The latter performed a one-dimensional hydrodynamic simulation based on the evolutionary tracks and the
simulations of the circumstellar shell around Asymptotic Giant
Branch (AGB) stars. The results of their simulation showed that a shock-bounded ionized main PN shell, moving supersonically into the
AGB-material, compresses the inner parts of the matter into a
dense but thin shell. The unaffected AGB-material becomes
ionized as well and forms a rapidly expanding halo. Because
of the drop in luminosity of the central star of the PN (CSPN)
towards the white dwarf regime, the outer part of the PN shell
recombines quickly forming a second inner halo. The disadvantage
of this simulation is that it just considers one dimension.
Thus, more complex interactions of the winds such as
Rayleigh-Taylor instabilities, which quickly build a clumpy environment
as we see in HST images, are not taken into account.
For this reason, the filling factor ($\epsilon$) has been usually set to unity.
Moreover, comparing these models with observations of older evolved nebulae suffers from interaction to the environment and the interstellar matter (ISM).
Thus, the study of nebulae with higher galactic latitudes is of special interest. However, there might be an influence on other properties changing the wind physics, mainly due to the lower metallicity. Often photoionizing radiative transfer codes like the steadily evolving  Cloudy\footnote{\url{http://www.nublado.org}} \citep{Cloudy90,Cloudy94,Cloudy13,Cloudy17} are therefore used to characterize such nebulae and to identify their membership to the stellar populations \citep{Em04,Em05,oettl14}.\\
The PN IC~5148 (PN~G002.7-52.4) is a nebula which is to date not well investigated in detail. First listed in the Second Index Catalogue of Nebulae and Clusters of Stars \citep{ICcat} with two independent entries as numbers IC~5148 and IC~5150 discovered by \citet{DiscoverySwift} and \citet{DiscoveryGale} independently, it was finally discovered to be the same object by \citet{HO61}.
Morphological it is declared as a round nebula in all catalogues and \citet{MSPN1} classified it as a multiple shell planetary nebula (MSPN) due to a small step in the \ion{H}{$\alpha$} image with a radius ratio of only 1:1.2 between the two structures. In addition, the intensity decrease was found smaller than that for typical MSPNe. A few years later the authors searched with a larger field systematically around many nebulae for extended emission features without detecting the very low surface brightness halo we investigate here \citep{MSPN2}.
While earlier spectroscopic studies took only a small fraction of one or two spectral lines to obtain radial velocity and expansion of the nebula \citep[e.g.][]{Velocity1988}, to our knowledge up to now the only, more extended spectroscopic analysis were performed by \citet{KSK90} and by \citet{KB94} within two surveys of 75 and 80 southern PNe, respectively. Both studies used pre-CCD electronic spectral scanning devices, taking a very small aperture region of the nebula and detected only a hand full of lines. They end up coinciding in the result that the nebula has about galactic disk abundance or only slight underabundance despite its large galactic latitude ($b\sim-52\degr$).
Further, the survey spectra used in the PN catalogue of \citet{Acker92} gives only intensities of 4 lines. The inspection of the original data file of this survey, provided now at the Hong Kong/AAO/Strasbourg PN data base \citep[thereafter HASH\footnote{\url{http://hashpn.space/}};][]{HASH1,HASH2}, does not recover more usable lines above the noise level.\\
The CSPN of IC~5148 was first classified by \citet{AC82}, and studied in detail by using the International Ultraviolet Explorer (IUE) by \citet{KF85}. They revealed that it is a very evolved hot central star. The IUE color index was not sufficient to derive an exact value of the temperature, but they give values based on their 3 methods from approximately 100 to 220\,kK. As later stated by \citet{PP91a,PP91b}, using the same data, no strong wind features were found. Moreover, a latter investigation shows that this is common for CSPN of nebulae which have passed the knee in the Hertzsprung-Russell diagram (HRD) towards the beginning white dwarf cooling track. The only published optical study of the CSPN is the classification of \citet{Mendez91}, giving it the type of a hgO(H) star (''{\sl a hydrogen rich high gravity star with very broad Balmer absorptions}'').\\
Previous distance estimations obtained from so called statistical distance scales span from 0.53\,kpc using the CSPN evolution \citep{Phillips05},
1.06\,kpc using the ionized mass versus surface brightness relation \citep{CKS92,SSV08}, 1.1\,kpc using the IRAS dust temperature \citep{IRAS_dist}, 1.37\,kpc by using the H$\alpha$ surface brightness-radius relation \citep{frew16}, 1.8\,kpc using a nonlinear calibration of the radio brightness vs diameter \citep[using ][calculated by us]{BS95}, 2.1\,kpc using the radio temperature vs radius relation  \citep[using ][calculated by us]{Steene_Zijlstra94} and even 3.59\,kpc using the mass-radius distance \citep{Zhang95}.\\
We present in this paper the recent discovery of extended emission features around the hitherto known nebula IC~5148 and a detailed medium-high resolution spectroscopic study of its central star with the ESO VLT X-SHOOTER instrument. We also performed a detailed spectroscopic study with long-slit spectroscopy from ESO FORS2 in various directions of the nebula and the newly discovered halo. The latter is to our knowledge one of the first and up to now deepest of its type in a halo.
Finally, using the now available distance estimation from the second data release DR2 of GAIA \citep{gaia_dr2} and the derived parameters from our CSPN spectroscopy, we modeled the main nebula by using the photoionizing radiative transfer code Cloudy \citep{Cloudy17}.

\begin{figure}[ht!]
\centerline{\includegraphics[width=60mm]{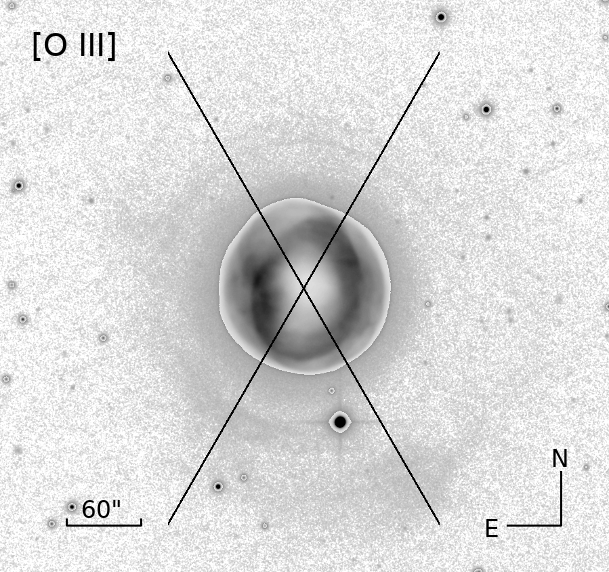}}
\centerline{\includegraphics[width=60mm]{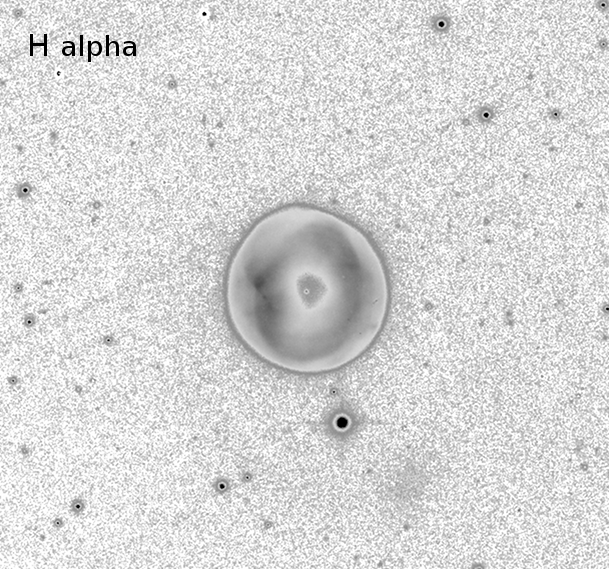}}
\caption{The coadded direct images in [\ion{O}{iii}] (upper) and \ion{H}{$\alpha$} (lower). The slit positions of the spectroscopy are indicated in the upper panel. While in the inner part a linear intensity scale covering the whole range from 0 to peak flux of the nebula was used, for regions with flux below 5\% a steep {\tt asinh} between 0.01 and 1\% of the peak of the linear part was used to enhance the by orders of magnitude fainter structures.}
\label{fig_direct}
\end{figure}

\section{Data}

\noindent
The narrow band imaging was derived at the CDK20 Planewave telescope of ITelescope.net in Siding springs (31\degr16\arcmin24\arcsec~south, 149\degr03\arcmin52\arcsec~east) during several nights in July 2013. An Apogee U16MD9 camera equipped with a Kodak 16803 CCD without focal reducer and without binning was used. A total exposure time of 11.0\,h with the [\ion{O}{iii}] filter by co-adding 22 frames and 4.5\,h with the \ion{H}{$\alpha$} filter by co-adding 9 exposures were obtained. The resulting imaging, shown at Fig.~\ref{fig_direct}, was used to identify the features of interest and the position angles for the follow up investigations.
To obtain a correct line ratio [\ion{O}{iii}]/\ion{H}{$\alpha$} image, the resulting frames were differentially calibrated by means of the long-slit spectroscopy derived at ESO. The astrometry was determined by 40 surrounding stars of the GAIA DR2 catalogue \citep{gaia_dr2} resulting in a plate scale of 0\farcs5447 per pixel. {Stars nearby the nebula on the coaded images have a FWHM of 1\farcs82 and 1\farcs96 in} \ion{H}{$\alpha$} {and} [\ion{O}{iii}] {respectively.}
\begin{figure*}[ht!]
\sidecaption
\includegraphics[width=120mm]{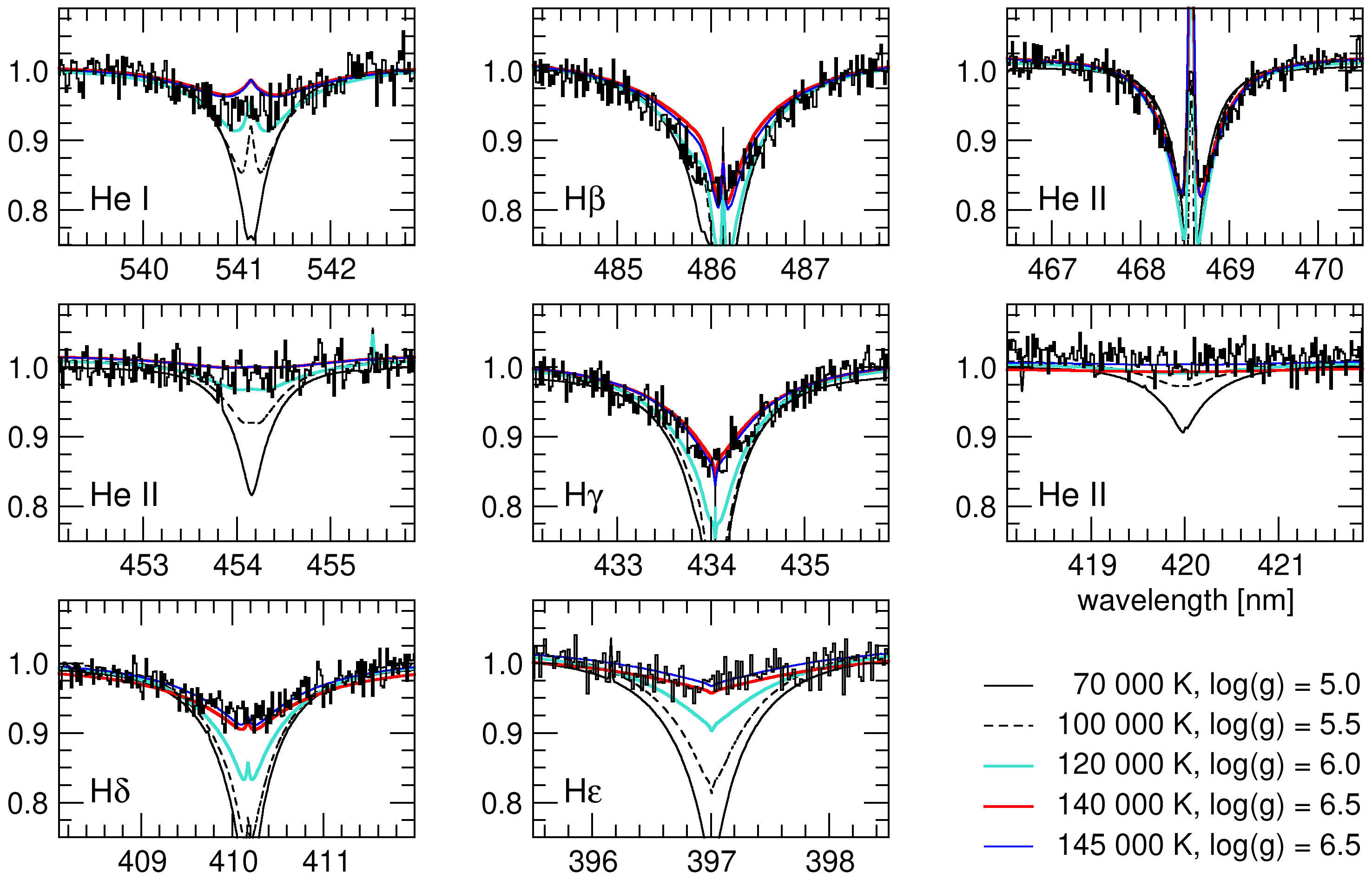}
\caption{Spectral lines from the medium-high resolution spectroscopy of the UVB Arm of X-SHOOTER. Overlaid are the model grid spectra along the evolutionary track $\log(g)$ values of a CSPN according to \citet{Miller_Bertolami} from the high resolution stellar spectra grid from the German Astrophysics Virtual Observatory (GAVO) grid resources \citep{GAVO2}. }
\label{fig_xsh_line}
\end{figure*}

\noindent
Long-slit spectra were obtained with the FORS2 spectrograph \citep{fors} mounted on the Cassegrain focus of ESO VLT UT1 (Antu) in the nights October 5$^{\rm th}$, 2016 from 3:15 to 4:03\,UT; October 6$^{\rm th}$, 2016 from 0:15 to 0:30\,UT and, October 10$^{\rm th}$, 2016 from 1:17 to 1:45\,UT in service mode as filler program with minor sky quality.
In total 7 spectra were collected, 3 with a position angle PA=30\degr~and 4 with a position angle PA=150\degr~(from N over E) on the sky through the center of the PN. The slit positions were selected to cover the features in the wide halo of the PN. We used the long-slit mode of FORS2 since this enabled us to cover the entire halo  using the full slit length of 6\farcm8 with the standard collimator (slit width = 0\farcs7). All spectra were taken with 14 minutes exposure time.
We used GRISM 300V and the GG435 order separation filter, covering a wavelength range from 455 to 889\,nm. The MIT/LL CCD mosaic and the standard focal reducer collimator result in a 0\farcs2518 pixel$^{-1}$ spatial scale. This setup leads to a final wavelength dispersion of 0.33\,nm\,pixel$^{-1}$. The night sky lines were measured with a resolution $R = \lambda/\Delta\lambda$ of 200 and $R = 360$ at the blue end and the red end of the spectrum, respectively. We derived a FWHM of the stellar sources along the slit of about 1\farcs17 at the blue end and 0\farcs95 at the red part of the spectrum. This corresponds well to the reported DIMM seeing of the ESO meteo monitor of 1\farcs1\,@500\,nm.
The data were reduced incorporating the standard calibration mode using the ESO FORS2 pipeline v5.3.11\footnote{\url{http://www.eso.org/sci/software/pipelines/fors/fors-pipe-recipes.html}} \citep{pipeline}.
The resulting flux calibration was compared with the expected continuum flux of the CSPN, finding a scatter less than 5\% between the individual observations obtained at clear, but non-photometric sky conditions. Two spectra at position angle 30\degr~were taken at slightly minor sky conditions {with some cirrus clouds} at October 5$^{\rm th}$. These were scaled with a constant wavelength independent factor using the central star observations. We used the software package {\sl molecfit} \citep{molecfit1,molecfit2} to create a model of the telluric absorption lines by applying it on the high S/N central star spectrum. As the PN and its extended halo covered the whole slit length, we used the software {\sl skycorr} \citep{skycorr} to correct for the sky emission lines. This approach also enabled us to derive a secondary correction of the wavelength calibration. We averaged the spectra for each sky direction to achieve the final spectra of the faint sources used for this study. Finally, a correction by photometric values of 20\% was applied (see next section).

\noindent The medium-high resolution echelle spectra were taken with the X-SHOOTER instrument \citep{Vernet2011} mounted at the ESO VLT UT2 (Kueyen) in the night October 5$^{\rm th}$, 2016 from 0:23 to 0:57\,UT at slightly degraded weather conditions as filler program in service mode. This instrument covers the entire wavelength range from 0.3 to 2.5 $\mu$m in three spectral
arms (UVB arm from the ultraviolet to the B band with an E2V CCD44-82; VIS arm in the visual regime with a MIT/LL CCD; NIR arm in the near-infrared regime with an Hawaii2RG chip), at medium resolution simultaneously. While the NIR and UVB frames had an exposure time of 125 seconds each, the VIS arm exposures were 213 seconds each (all 0.8\arcsec slit). A total of 29 frames were obtained and co-added during data reduction with the standard ESO X-SHOOTER pipeline v2.8\footnote{\url{https://www.eso.org/sci/software/pipelines/xshooter/xshooter-pipe-recipes.html}} \citep{x-shooter-pipeline}. The spectra were normalized to continuum to derive line profiles for the spectral analysis. Thus, the slightly degraded weather conditions did not harm. We did not apply a telluric absorption correction since all lines of interest were in undisturbed regions. The signal to noise ratio was about 45 in the UVB and 25 in the VIS arm. That was lower than expected from exposure time estimators and the DSS magnitude of the central star. It could be caused by a combination of cirrus clouds and the 1.6\arcsec\,\,seeing at that time. The signal to noise ratio of the NIR part of the spectrum was insufficient to derive any parameters from there and were not further used.

\section{Analysis}
\subsection{The central star}

\begin{figure}[t!]
\centerline{\includegraphics[width=88mm]{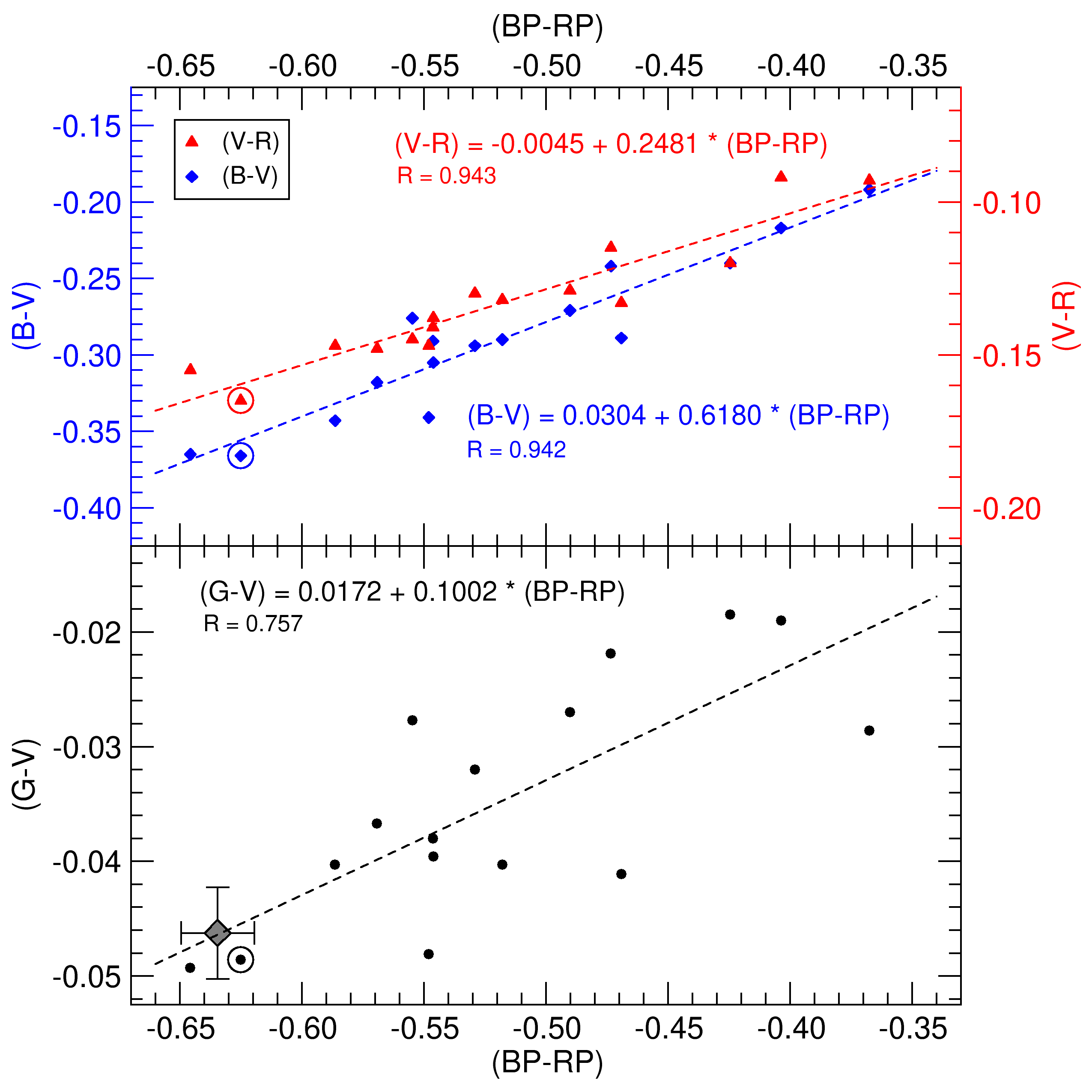}}
\caption{The color equations of hot white dwarfs from the Landolt photometric standards \citep{Landolt92} and from the spectrophotometric standards \citep{Landolt07} versus GAIA DR2. The target position used for calibrations is given in the lower panel. While all other stars are hot nearby compact white dwarfs without PNe, the encircled target is the central star of the Helix Nebula.}
\label{fig_gaia_phot}
\end{figure}

The X-SHOOTER spectrum was normalized to unity flux and searched for absorption lines. The only one appearing in the VIS arm spectrum was the \ion{H}{$\alpha$} line. Nevertheless, it is heavily contaminated by the emission of the Hydrogen line by the nebula and by the two strong [\ion{N}{ii}] lines. As shown below in the following section, these lines are asymmetric and doubled due to nebular expansion. Moreover, they vary significantly along the slit. Thus, a reliable removal was not possible and the \ion{H}{$\alpha$} line was not used for the further analysis. The NIR arm has a too low S/N ratio ($<$~10 in J band and $<$~5 in H\&K) to derive a useful spectrum. Therefore, the analysis of the CSPN parameters is focused only on the UVB arm of the X-SHOOTER instrument.
\begin{figure}[ht!]
\centerline{\includegraphics[width=88mm]{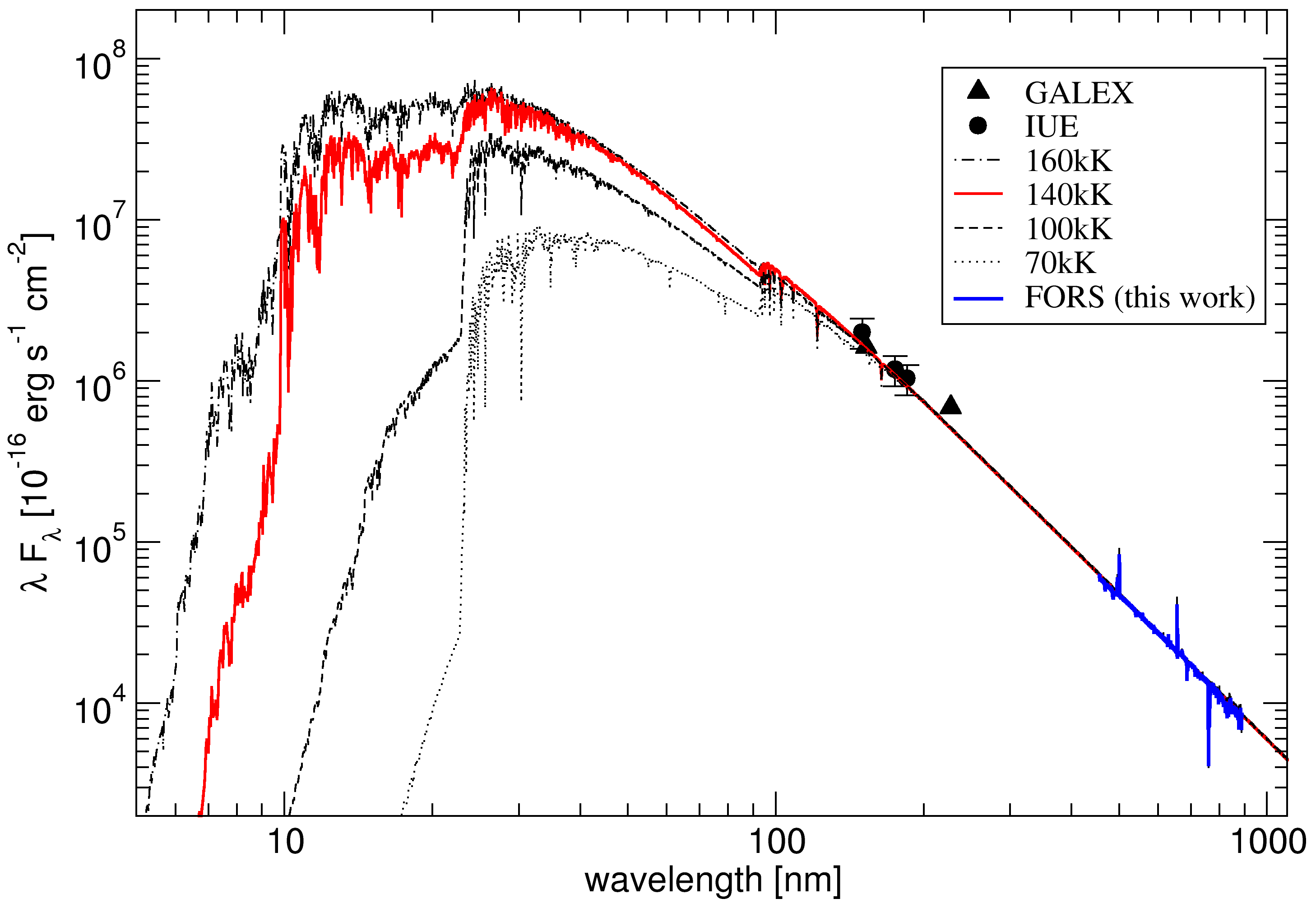}}
\caption{The UV photometry from IUE and GALEX with our calibrated FORS2 spectrum. The model spectral energy distributions from TMAP are overlaid, normalized to the FORS2 data.}
\label{fig_cspn_total}
\end{figure}
\\
We overlaid in our analysis the high resolution non-LTE spectra for central stars of planetary nebulae of the T\"ubingen stellar atmosphere \citep[thereafter TMAP,][]{TMAP} as offered at \citet{GAVO2} to the identified lines in the UVB arm spectra (Fig.~\ref{fig_xsh_line}). The clear appearance of the \ion{He}{ii} 468.6\,nm line in absorption, which was not seen by \citet{Mendez91}, and the \ion{He}{ii} 420.0+454.1\,nm lines without major absorption, places the CSPN in their classification as O(H) and not as hgO(H). The number of lines and the signal to noise does not allow us for a sophisticated determination of the $\log(g)$ value. Thus the evolution along a standard model by \citet{Miller_Bertolami}
was assumed. As \cite{Napiwotzki2001} showed, the result is not sensitive on the selection of $\log(g)$ for the optical flux analysis.
A likelihood analysis of all lines, weighted by equivalent width of the line to consider the weakness of the lines like \ion{He}{ii} 454.1\,nm, 420.0\,nm or \ion{H}{$\epsilon$}, was used to derive a $T_{\rm eff}\,\approx\, 140^{+5}_{-13}\,{\rm kK}$ ($\log$ T = 5.146) for the central star.\\
The calibrated FORS2 spectra of the CSPN were first compared individually 
and then combined to derive the final flux.
The derived flux curve was folded with the filter curves of \citet{Bessel1990} and multiplied with the zero point given in \citet{Bessel1998}\footnote{There in \citet{Bessel1998} Tab. A.2 the $zp(f_\lambda)$ and the $zp(f_\nu)$ are reversed in the formulas for the magnitudes - confirmed by author in private communication.}. This results in V=16\fm34.
The DENIS I$_{\rm C}=16\fm476\pm0\fm10$ lead with a typical color of such very hot white dwarfs (V-I$_{\rm C})\approx -0\fm35\pm0\fm05$ from \citet{Landolt07} for a V$_{\rm DENIS}=16\fm13\pm0\fm15$, but the target is very much near the survey limit. \citet{KF85} derived magnitudes of V(UV)\,=\,16\fm16$\pm$0\fm17 using the IUE spectra between 130 and 185\,nm and assuming a 130\,000\,K blackbody spectrum and zero extinction due to the high galactic latitude.
The flux calibration allow us directly to estimate the interstellar foreground extinction to be very low. The Balmer decrement of the nebular lines in our FORS2 spectra (only \ion{H}{$\alpha$} and \ion{H}{$\beta$} are covered by these spectra) give us an E(B-V)=0\fm031$\pm$0.010. The central star continuum, using stellar models from the T{\"u}bingen Model Atmosphere Package TMAP \cite{TMAP} accessed via the German Astrophysical Virtual Observatory (GAVO) service TheoSSA \citep{GAVO1}, confirm the nearly extinction free line of sight. The Galaxy Evolution Explorer (GALEX) data base give the similar value of E(B-V)=0\fm021 based on the NUV flux of $m_{\rm NUV}=13\fm480\pm0\fm004$.
Using the GALEX FUV data in the small 7\arcsec aperture and applying that to TMAP NLTE model together to the effective temperature derived by our spectroscopy with X-SHOOTER and, the interstellar extinction obtained from the Balmer lines, with the extrapolation as it is used in \citet{KF85} for IUE data, we achieve for a visual magnitude of 16\fm35. The spectral energy distribution (SED) from 130 to 800\,nm fits very well using a low extinction value of E(B-V)=0\fm03 (see Fig.\ \ref{fig_cspn_total}). The overlaid TMAP spectra also show, how marginally the SED differ for these stars even at IUE and GALEX wavelengths, and thus why \citet{KF85} were unable to derive a more accurate CSPN temperature.
We further adopt this low value for the interstellar extinction. This value is in contradiction to the spectroscopy by \citet{KSK90} and \citet{KB94}, who found, using several lines of the Balmer decrement, values as high as $c($\ion{H}{$\beta$}) of 0.4 and 0.38 respectively (corresponding to E(B-V)=0\fm26). However, using their published line intensities from \ion{H}{$\alpha$}, \ion{H}{$\beta$}, \ion{H}{$\gamma$} and \ion{H}{$\delta$} individually give the large scatter of values of $c$ from even slightly negative values of -0.03 up to 0.85. An extinction value as high as $c = 0.4$ does neither match with our spectra, nor with the IUE and GALEX photometry. Moreover, it is also unlikely with respect to the high galactic latitude of the target\footnote{{https://ned.ipac.caltech.edu/?q=extinction\_calculator} provided by the NED data base and based on \citet{SF11}}. On the other hand the \ion{H}{$\alpha$}/\ion{H}{$\beta$} = 1.7 in \citep{Acker92} is nonphysical. We thus do not take these values further into account.\\
GAIA DR2 has some issues with extraction of nebular contamination \citep[see][]{GAIA_PN18}. However, since the central star of IC~5148 appears isolated from its main nebula, we expect for a mostly reliable photometry from GAIA. We thus compared isolated blue compact white dwarfs in the lists of Landolt photometric standards \citep{Landolt92} and spectrophotometric standards \citep{Landolt07}
to derive color equations dedicated to this type of stars for the photometry into standard system (see Fig.\ \ref{fig_gaia_phot}).
With this we derive a V$_{\rm GAIA}=16\fm14\pm0\fm03$ {, using G$_{\rm GAIA}=16\fm097$ and GAIA color (BP-RP)=-0.635.}
We finally adopted the latter value as reliable, and corrected the FORS2 fluxes by 20\% to take into account the filler program minor sky conditions.
\begin{figure}[t!]
\centerline{\includegraphics[width=88mm]{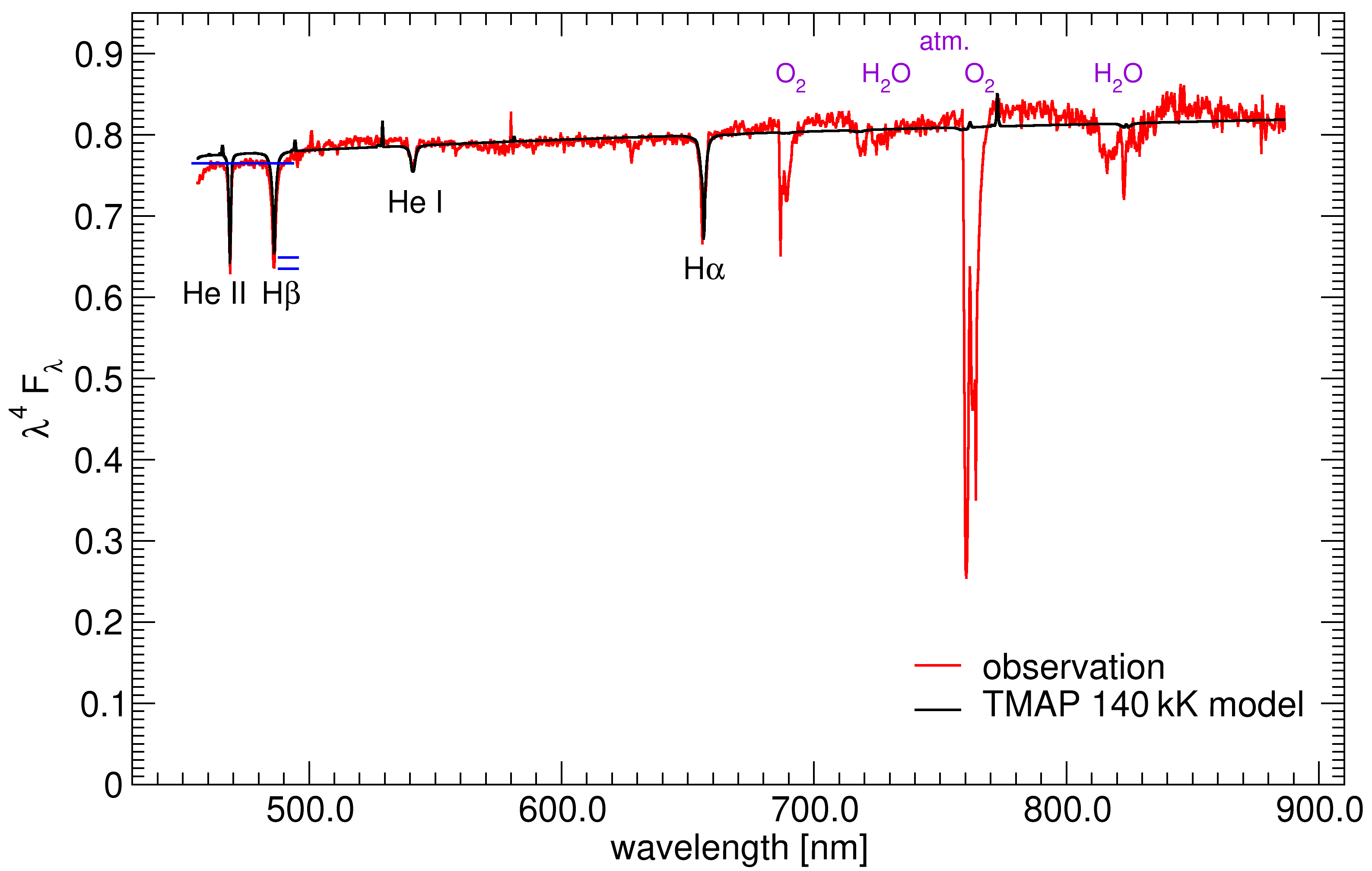}}
\caption{The $\lambda^4$ weighted FORS2 spectrum with the TMAP model. We intentionally did not remove the telluric O2 and H$_2$O absorptions at the red end of the spectrum. The blue marks show the required correction of the line depths at the blue end due to the deviation to the continuum.}
\label{fig_fors_flat}
\end{figure}
\\
To test the result and to approve for the photometric flux calibration, the FORS2 spectrum was weighted with $\lambda^4$ (giving for the Raleigh Jeans tail of an exact blackbody a flat spectrum) and compared with the TMAP model downgraded to the FORS2 resolution (see Fig.\,\ref{fig_fors_flat}). The deviation of the continuum is in most parts of the order of a few percent. The four stellar lines in the region, namely H$\alpha$, H$\beta$, He I 540.6\,nm and He II 468.5\,nm (taking into account the slight error in the continuum calibration at the blue end) fit also perfectly in their depths.  The latter together to the fact that the He~I/He~II ratio is very sensitive to the stellar temperature, independently confirms our X-SHOOTER result on $T_{\rm eff}$.

\subsection{Nebular expansion, systemic velocity and astrometry}
\label{sec_nebular_expansion}
\begin{figure}[t]
\centerline{\includegraphics[width=85mm]{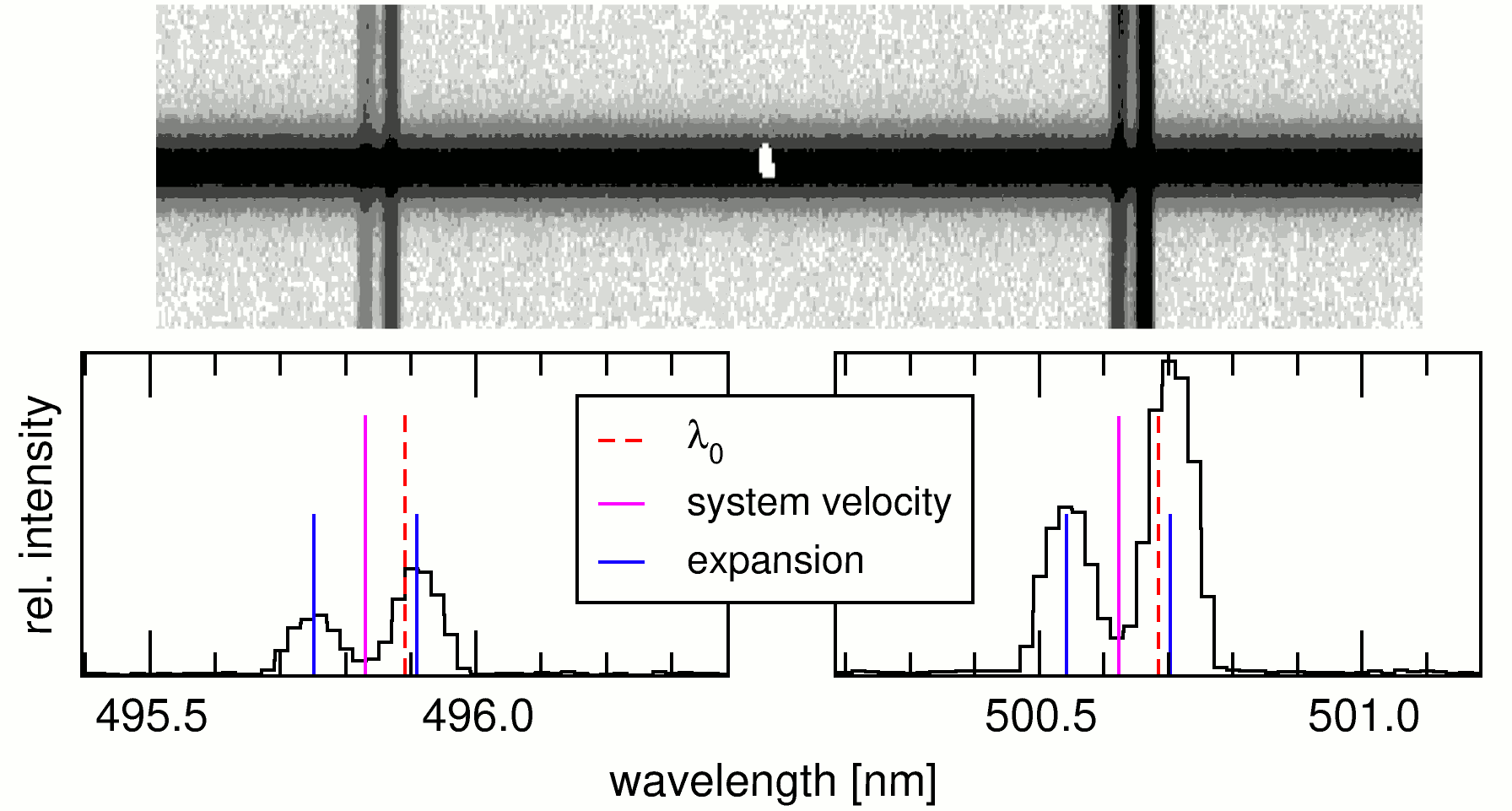}}
\caption{Expansion of the PN and system velocity in the Galaxy from X-SHOOTER spectra after correction for LSRK. }
\label{fig_expansion}
\end{figure}

The X-SHOOTER spectra have only a 9\arcsec~long-slit and thus do not cover the nebula as a whole. But its high resolution allows the determination of the line splitting along the line of sight in the center of the nebula, giving thus the direct value for the nebular expansion velocity (see Fig.~\ref{fig_expansion}). Moreover, the midpoint of the two lines gives a good value for the systemic velocity. Low resolution spectroscopy often suffer from the fact that the two unresolved lobes are not identical in intensity. Thus for example in the case of unresolved lines the result would be biased towards to red values. The measured expansions were independent of the element type within the error.
This is in fact astonishing, as often the high ionized species give higher expansion velocities than the low ionized species \citep{Weinberger,Gesicki2014}.
We measured a peak to peak expansion velocity of 47.9$\pm$1.5\,km\,s$^{-1}$. This is in very good agreement with the value of 53.4\,km\,s$^{-1}$ given by \cite{Velocity1988} if taking into account the much lower spectral resolution of 11.5\,km\,s$^{-1}$ used by the authors. The nebula thus has a fairly large expansion compared to other roundish nebula in the catalogues \citep[e.g.][and citations therein]{Weinberger}.
The center between the two well separated peaks was used as system velocity.  The measured system blueshift was corrected for the local standard of rest (LSRK) using the IRAF\footnote{Image Reduction and Analysis Facility available at http://iraf.noao.edu/} procedure {\em rvcorrect}. This result in $RV_{\rm LSRK} = -36.7\pm1.7$\,km\,s$^{-1}$. The systematic difference to the hitherto published value of -23\,km\,s$^{-1}$ \citep{Velocity1988} most likely originates from the asymmetry of intensities of the redshifted and of the blueshifted part of the shell. This moves the center towards the brighter peak in the case of low resolution spectroscopy.
\\
The astrometry was derived by 40 surrounding stars from the GAIA DR2 catalogue \citep{gaia_dr2} which resulted in a rms of about 105\,mas. We found only one old plate, which was scanned with high resolution from the ESO red survey plate scanned at MAMA and provided via CDS Aladin. It dates from September, 5$^{\rm th}$ 1985. This was calibrated in the same homogeneous way with the same GAIA stars. The rms of this calibration was 80\,mas.
The resulting positions using our astrometric solution and that of GAIA \citep{GAIA_DR2_ASTROMETRY} are given by:

\medskip
\noindent \begin{tabular}{rlll}
 \multicolumn{4}{l}{GAIA DR2: Epoch 2015.5, Equinox J2000.0}\\
  $\alpha$ &=& $\phantom{-}21^{\rm h}59^{\rm m}35\fs095392053$ & $\pm 0\fs000000001$ \\
  $\mu_\alpha$ &=&  $-0.00652\,{\rm arcsec\,yr}^{-1}$ \\
  $\delta$ &=& $-39\degr23\arcmin\,\,08\farcs35833212$   & $\pm 0\farcs00000002$ \\
  $\mu_\delta$ & = & $-0\farcs00779\,{\rm arcsec\,yr}^{-1}$\\
%
 \multicolumn{4}{l}{CDK20: Epoch 2013.52, Equinox J2000.0}\\
  $\alpha$ &=& $\phantom{-}21^{\rm h}59^{\rm m}35\fs1066$ & $\pm 0\fs0080$ \\
  $\delta$ &=& $-39\degr23\arcmin\,\,08\farcs439$   & $\pm 0\farcs12$ \\
 \multicolumn{4}{l}{ESO R: Epoch 1985.68.5, Equinox J2000.0}\\
  $\alpha$ &=& $\phantom{-}21^{\rm h}59^{\rm m}35\fs1084$ & $\pm 0\fs0053$ \\
  $\delta$ &=& $-39\degr23\arcmin\,\,08\farcs714$   & $\pm 0\farcs08$ \\
\end{tabular}
\medskip

\noindent While the ESO R plate corresponds perfectly using the GAIA proper motion, the stacked CDK20 images have a offset slightly bigger than the error bars, most likely having its origin in a remaining distortion correction on the stacked images. But as it is still far below the FWHM caused by the seeing, we are still able to trust the division of images to derive line ratios. Using the GAIA DR2 distance of 1.3\,kpc {(see section \ref{sec_distance})} we derive a proper motion velocity in the plane of sight of $62\,\pm5\,{\rm km}\,\,{\rm s}^{-1}$ towards southeast.

\subsection{Distance, luminosity and position in the Galaxy}
\label{sec_distance}
The most commonly used statistical distance scales are that one of \cite{CKS92} and \citet{SSV08} giving a distance to IC~5148 of 1.06\,kpc. Using the same radio data, but the radio brightness temperature method of \cite{Steene_Zijlstra94} we obtain a distance of 2.1\,kpc. More recently, \citet{frew16} established a method based on the optical H$\alpha$ surface brightness giving 1.37\,kpc.
Using a recent calibration with the GAIA DR1 data for the statistical methods mentioned above by \citet{Stanghellini2017}, we then get a distance of 1.88$^{+0.53}_{-0.48}$\,kpc. In a newly investigation, containing a large sample of sources, \citet{GAIA_PN18} showed for some issues concerning GAIA DR2 \citep{gaia_dr2,GAIA_DR2_PARALAXES} parallaxes for PNe with high surface brightness around the central stars. However, in this case the blue GAIA color (BP-RP) suggests for no contamination, and so we believe we are able to use the value provided by GAIA. Thus, we get a GAIA distance of $D_{\rm GAIA}=1.28^{+0.16}_{-0.13}\,{\rm kpc}$. For the further analysis a final distance of 1.3\,kpc is applied.
\begin{figure}[t]
\centerline{\includegraphics[width=88mm]{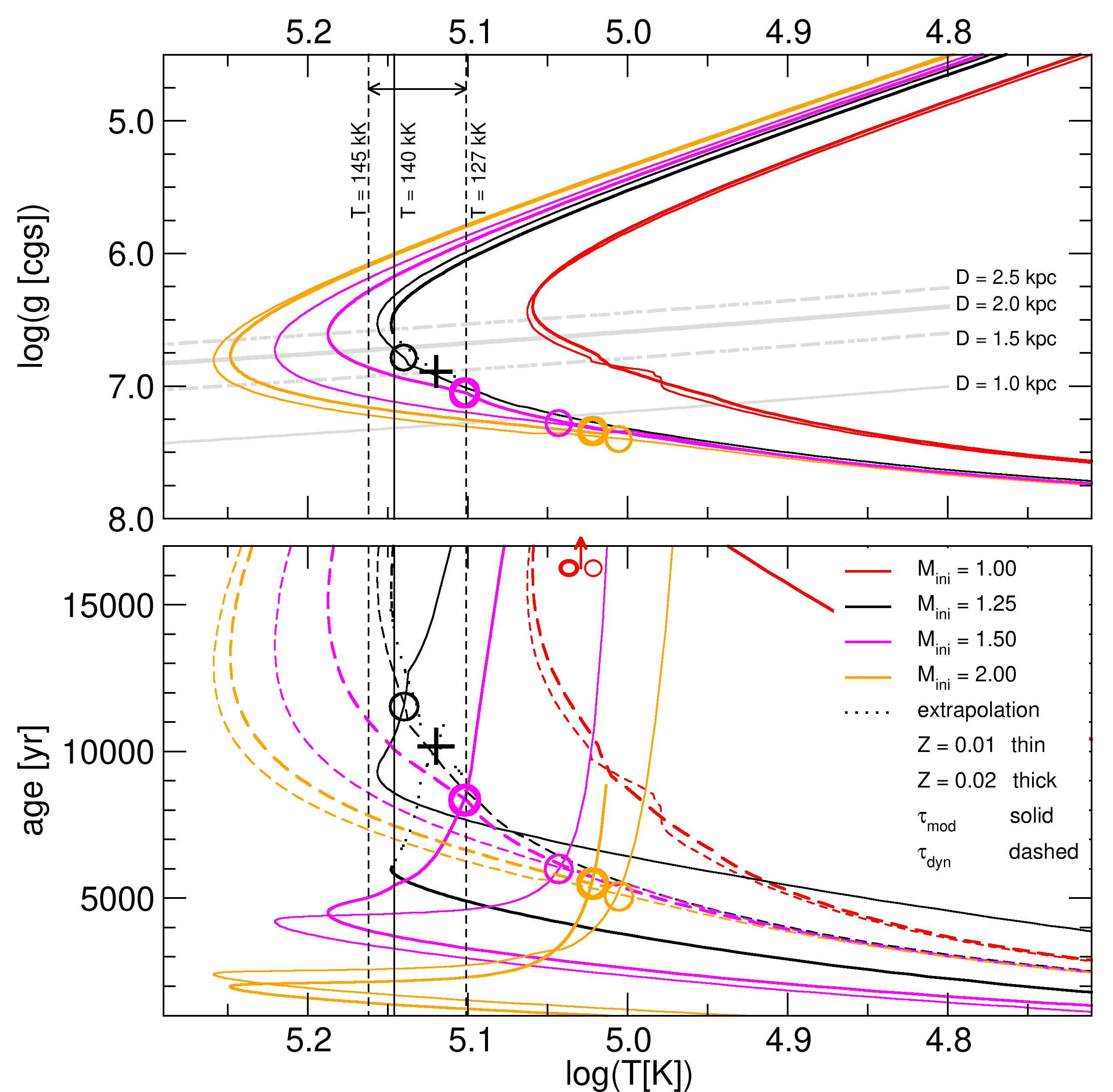}}
\caption{Upper: evolutionary tracks of \cite{Miller_Bertolami} as $\log(g)$ vs. $ \log(T)$ for initial masses of 1.00, 1.25, 1.5 and 2.0\,M$_\odot$ (color coded) for metallicities of Z=0.01 and Z=0.02 (line thickness coded). The corresponding spectroscopic distances are indicated in light grey. The vertical dark grey lines indicate the range of temperatures as derived from our spectroscopy. Dotted lines mark extrapolation of the track, which ended earlier. Lower: the computed model ages from the tracks above (solid lines) vs. the dynamical ages (dashed lines). The circles indicate the solutions by intersections of the model age with the dynamical age. The corresponding solutions found in lower panel are marked in the upper panel as well. Crosses mark the estimated solution of the extrapolated track. The intersections of the 1.0M$_\odot$ (red) fall far outside the graph.}
\label{fig_tracks}
\end{figure}
\\
The radial velocity, accurate proper motion and the GAIA distance allows us to derive the spatial movement vector. The target moves in galactic coordinates towards lower galactic longitude at 22\,km\,s$^{-1}$ and marginally increases its distance below the galactic plane with $-$7\,km\,s$^{-1}$ while it moves to larger galactocentric distance with $+$68\,km\,s$^{-1}$. Thus, while the peculiar fraction from normal galactic rotation and change in $z$ are small, it is significantly moving outwards in the Galaxy.\\
Using our spectra and photometry of the central star we are also able to derive a spectroscopic distance \citep{Napiwotzki1999,Napiwotzki2001} by means of the dereddened apparent magnitude, the mass of the central star and, the surface gravity $\log(g)$.
With a temperature derived from the X-SHOOTER data we obtained a range of distances. We compared these to the recent evolutionary tracks (Fig.\,\ref{fig_tracks}) by \cite{Miller_Bertolami}. We derived with the evolutionary tracks, the spectroscopic model distances and with the expansion velocity, a dynamical age $\tau_{\rm dyn}$ for each, {according to:}

$$ \tau_{\rm dyn} = \theta\,\,v_{\rm exp}\,\,D_{\rm phot}(T,M_{\rm fin},V_{\rm GAIA})$$
{where $\theta = 67\farcs2$ is the angular radius averaged over all directions, $v_{\rm exp} = 47.9 {\rm km s}^{-1}$ the expansion velocity, and $D_{\rm phot}$ is a distance as function of the effective temperature $T$ \citep[formula 1 in][]{Napiwotzki2001} using the final mass $M_{\rm fin}$ of the track and the photometry $V_{\rm GAIA}$.} This was then compared to the model age $\tau_{\rm mod}$. Even if we take into account an uncertainty of a factor of 2 in the dynamical age, the intersection within the given temperature range derived for the CSPN fits only for the following tracks: initial mass $M_{ini} = 1.25\,{\rm M}_\odot$, metallicity $Z = 0.01$ giving a distance of $D = 1.8\,{\rm kpc}$ and, $M_{ini} = 1.5\,{\rm M}_\odot$, metallicity $Z = 0.02$ ending up in $D = 1.3\,{\rm kpc}$. Extrapolating the track for $M_{ini} = 1.25\,{\rm M}_\odot$, metallicity $Z = 0.02$ (which stopped due to a late helium flash in the calculation of the stellar model grid) we derive an intermediate position at $D = 1.5\,{\rm kpc}$.
We obtain about the same results within 20\% using the tracks of \cite{VW94} and assuming an initial to final mass relation \citep{VW93} to derive mass and $\log(g)$. The tracks of \cite{Bloecker1995} do not cover such low mass stars. This intersection lowers slightly, but within the error by 0.8\,$\sigma$ the effective temperature of the star as derived above from the X-SHOOTER spectra.
Any other intersection of dynamical age with model age do not fit by more than a factor of three.
Hence, these evolutionary tracks are in good agreement with the distances and temperatures previously derived from the nebular parameters and lead us to the input parameter set for the best fitting model with a initial mass of $M_{ini}=1.5\,{\rm M}_\odot$ and Z\,=\,0.02. The resulting parameters are listed in Table\ \ref{tab_cspn}.

\begin{table}[t]
\centering
\caption{Derived input parameters for the CSPN.}
\label{tab_cspn}
\begin{tabular}{l l l}
\hline\hline
Parameter & symbol & value\\
\hline
    effective temperature & $T_{\rm eff}$ & $130\,{\rm kK}$ \\
                          & $\log(T_{\rm eff})$ & 5.114   \\
    initial mass & $M_{\rm ini}/{\rm M}_\odot$ & 1.50   \\
    final mass & $M_{\rm fin}/{\rm M}_\odot$ & 0.576  \\
    luminosity & $L/{\rm L}_\odot$ & 320    \\
               & $\log(L/L_\odot)$ & 2.51    \\
    surface gravity & $\log(g[{\rm cgs}])$ & 7.05 \\
    model age & $\tau_{\rm mod}$ & 8\,500\,years    \\
    \\
    apparent brightness & $m_{\rm V}$ & \phantom{$-$}116\fm14 \\
    distance & $D$ & \phantom{$-$}1.3\,kpc \\
    Galactic plane distance & $z$ & $-$1.0\,kpc \\
\hline
\end{tabular}
\end{table}

\subsection{Line nebular analysis}
\label{sec_nebular_analysis}
%
%
%
\begin{figure}[ht]
\centerline{\includegraphics[width=88mm]{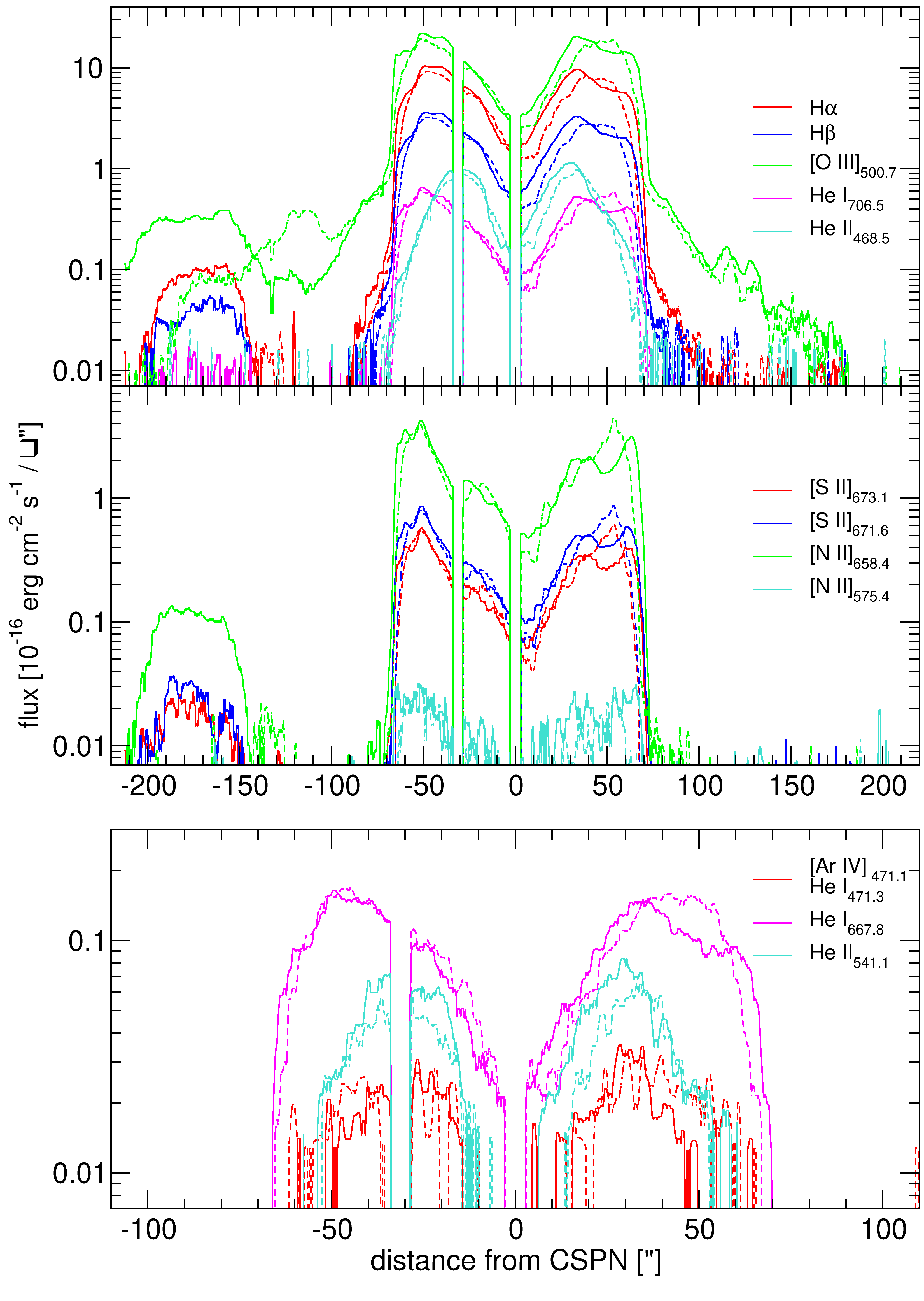}}
\caption{The spatial distribution of the major spectral lines. Upper: the brightest lines of oxygen, hydrogen and helium. Middle: the low ionized states of nitrogen and sulphur. The [\ion{N}{ii}] 654.8\,nm line was skipped for clarity as it caries the same information as the [\ion{N}{ii}] 658.4\,nm line. Lower: The fainter lines of the two different ionization stages of helium. As the line around 471\,nm shows the shape of the higher ionized \ion{He}{ii} lines, we have to assume that it is dominated by the [\ion{Ar}{iv}] 471.1\,nm emission and fewer contribution of the \ion{He}{i} 471.3\,nm emission. In all panels, solid lines denote to the slit at position angle 30\degr, while the dashed lines give the slit at 150\degr. The region contaminated by the central star was eliminated. Around 30'' south of the nebula center, the FORS2 instrument has a chip gap.}
\label{fig_lines}
\end{figure}

\begin{figure}[ht]
\centerline{\includegraphics[width=88mm]{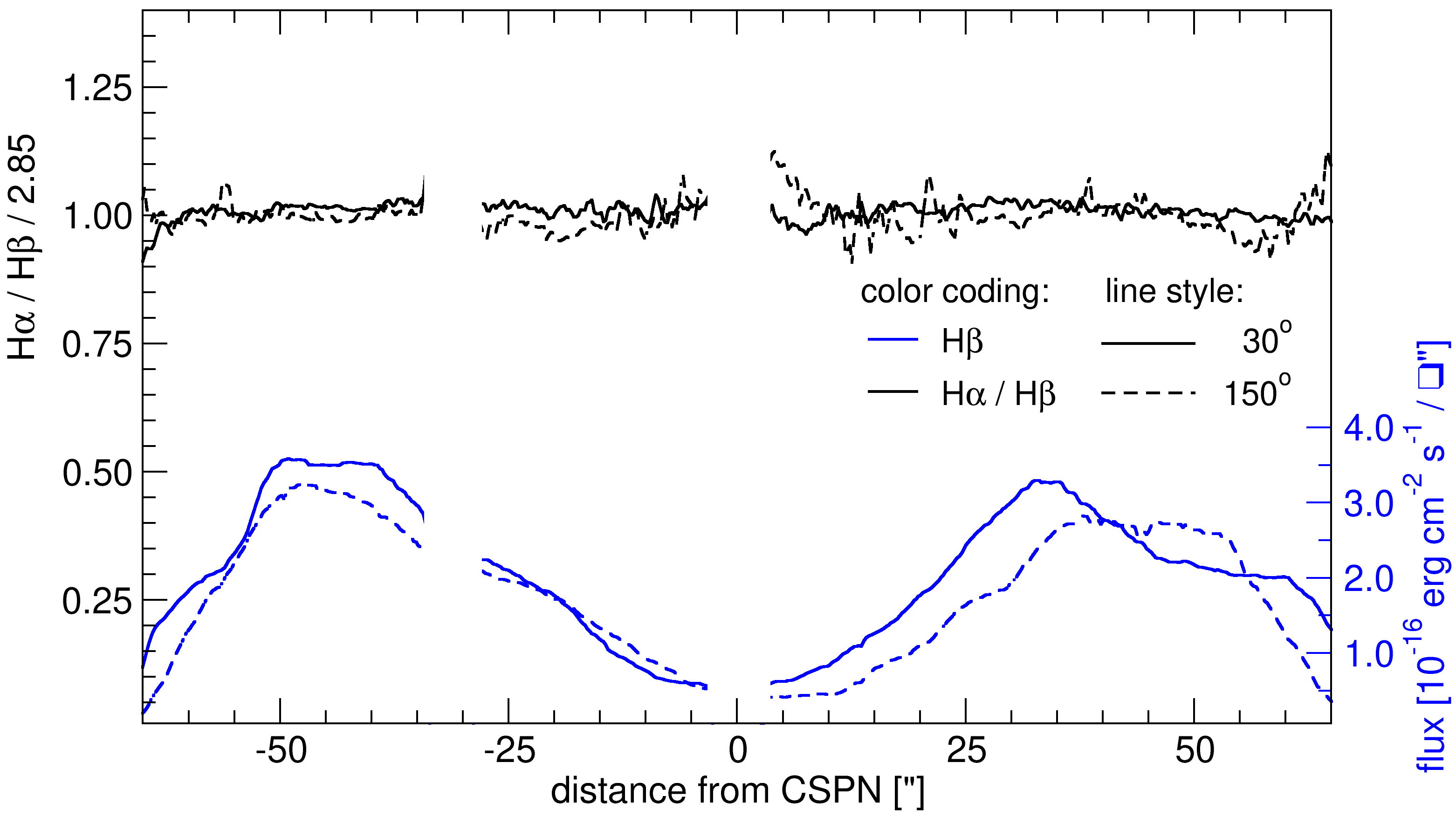}}
\caption{The \ion{H}{$\alpha$}/\ion{H}{$\beta$} normalized to a Case B line ratio \citep{OF06} to test for internal extinction.}
\label{fig_ext}
\end{figure}

\begin{figure}[ht]
\centerline{\includegraphics[width=88mm]{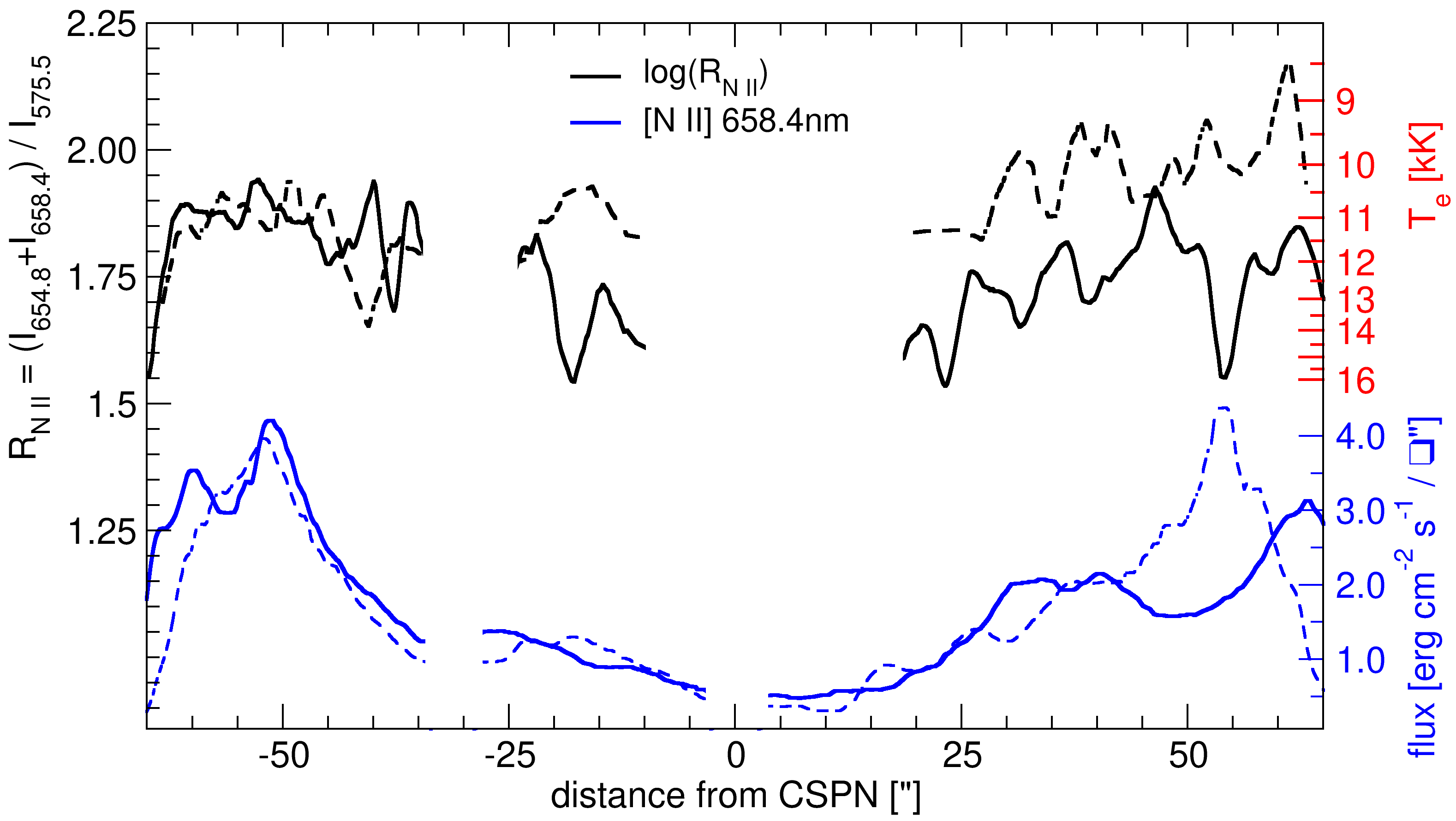}}
\caption{The electron temperature derived from [\ion{N}{ii}] lines. Despite the large variation in intensity itself, the temperatures are fairly homogeneous. Line coding for the two slit directions is as in Fig.~\ref{fig_ext}.}
\label{fig_Te}
\end{figure}

While the lines of \ion{H}{$\alpha$}, \ion{H}{$\beta$}, [\ion{O}{iii}] and \ion{He}{i} follow well the same spatial behavior of the main nebula rim (Fig.\ \ref{fig_lines}, upper panel), there is a clear concentration of the low ionized species [\ion{N}{ii}] and [\ion{S}{ii}] towards the outer edge (Fig.\ \ref{fig_lines}, middle). Moreover, more pronounced is the central concentration of the high ionized species \ion{He}{ii} and [\ion{Ar}{iv}] relative to the lower ionization lines of \ion{He}{i} (Fig.\ \ref{fig_lines}, lower panel).
The four cuts given by the two slits from the central star outwards in radial distance show fairly the same shapes, hence the lines were averaged as radial profiles for the Cloudy modelling below. The line ratio [\ion{O}{iii}] 500.7\,nm/\ion{H}{$\beta$} appears constant at high values of 7 to 8. On the other hand, the line ratio $R_{\rm He}$\,= (\ion{He}{ii}\,468.6\,nm/\ion{He}{i}\,587.5\,nm) which is normally used to define for an excitation class of the nebula as a total \citep{Ratag97,Martins02}, appears flat at $\approx$3.1 for an inner radius of 30'' and declines near perfectly linear to 0.5 out at 65''.
We derived the \ion{H}{$\alpha$}/\ion{H}{$\beta$} ratio for both slit positions to test for a significant fraction or variations of the internal extinction within the nebula. Despite the differences in line intensities and some fluctuations between the two slit positions, we were unable to detect any of such variations (Fig.\ \ref{fig_ext}).
\\
The $R_{\rm N\,II}=$([\ion{N}{ii}]\,654.8\,nm+[\ion{N}{ii}]\,658.4\,nm)/[\ion{N}{ii}]\,575.5\,nm line ratio was used to derive an electron temperature as we shown in Fig.\ \ref{fig_Te}. The blue nitrogen line has a fairly low signal to noise ratio, therefore the individual features observed at the line ratio should not be over-interpreted as intrinsic electron temperature variations, otherwise an spatial smoothing would be required. The temperatures lead to an average value of around $T_{\rm e}=11\,500$\,K using the calibrations given in \citet{OF06}. There is a slight tendency of an asymmetry in the northern part of the nebula, but in general terms it appears homogeneous with an electron temperature well above $10^4$\,K. Unfortunately there were no other temperature sensitive line triples like e.g. [\ion{O}{iii}], [\ion{Ar}{iii}] and [\ion{S}{iii}] covered in our spectral range. Temperatures derived from higher ionized states outside the colder clumps would have been of certain interest (see shock analysis section below).
\begin{figure}[ht]
\centerline{\includegraphics[width=88mm]{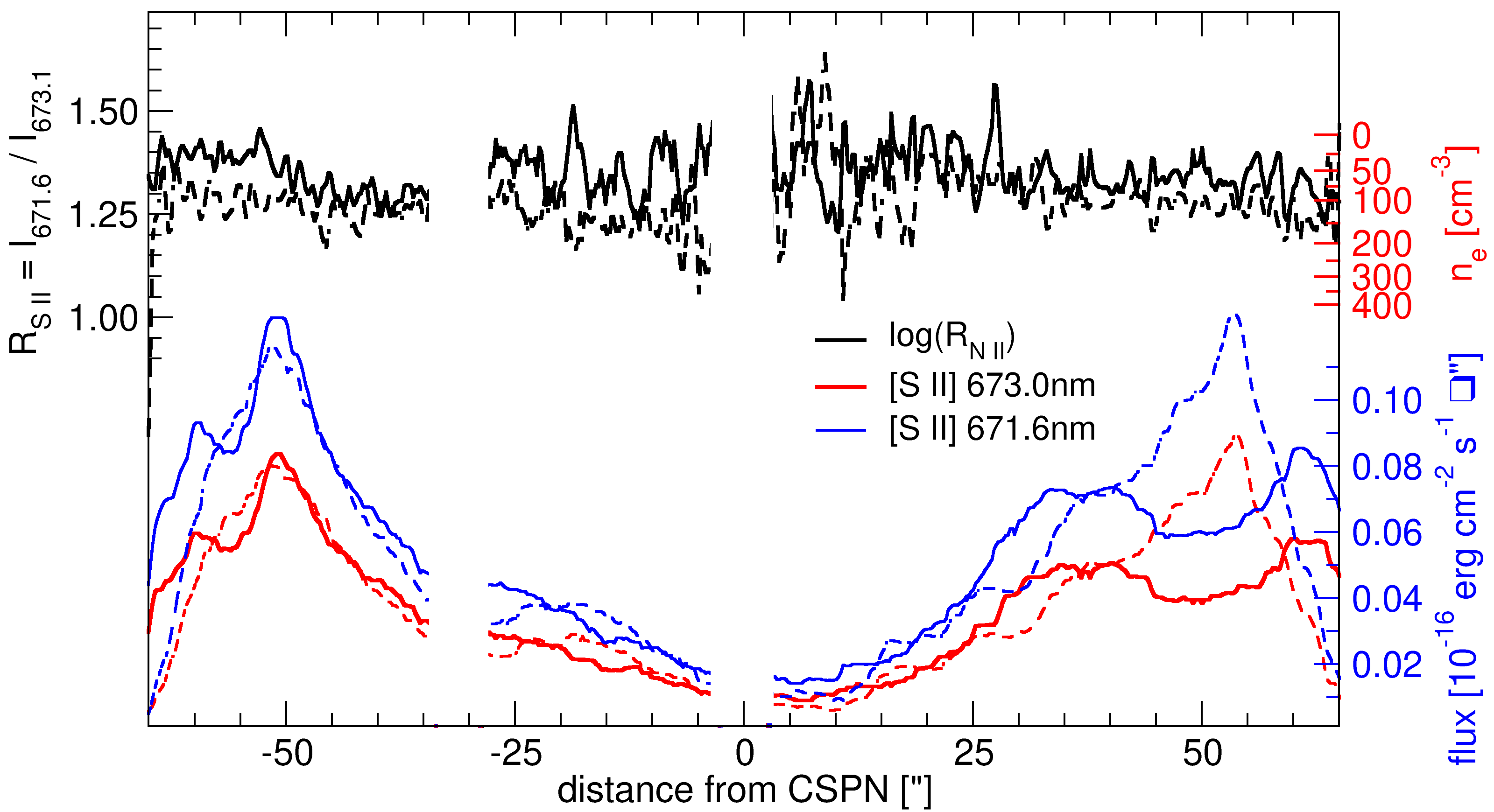}}
\caption{The density sensitive line ratio of [\ion{S}{ii}] lines. Despite the large variation in intensity itself, the temperatures are fairly homogeneous. Line coding for the two slit directions is as in Fig.~\ref{fig_ext}.}
\label{fig_rho}
\end{figure}
\\
Further we used the $R_{\rm
S\,II}=($[\ion{S}{ii}]\,671.6\,nm/[\ion{S}{ii}]\,673.1\,nm) line ratio to derive the electron density $n_{\rm e}$ in the nebula. We used the average electron temperature $T_{\rm e}$ derived above and the two calibration $\varrho(R_{\rm S\,II},T_{\rm e})$ provided by \citet{proxauf14}. The values are very constant at the lowest limit of the diagnostic diagram as shown in Fig.~\ref{fig_rho}. Thus the values have to be interpreted as upper limits mostly. As the flux varies significantly and thus the integrated volume emissivity which is a function of the average density <$\varrho$>, we have to assume taking into account this nearly constant values for $\varrho$, that the filling factor varies over the nebula. Something which was observed in the NIR molecular analysis for the nearby Helix nebula \citep{matsuura09} and was also a result of the modelling of NGC~2438 in \citet{oettl14}.

\subsection{Gas ionization tendency and shock heating}
\begin{figure}[t]
\centerline{\includegraphics[width=88mm]{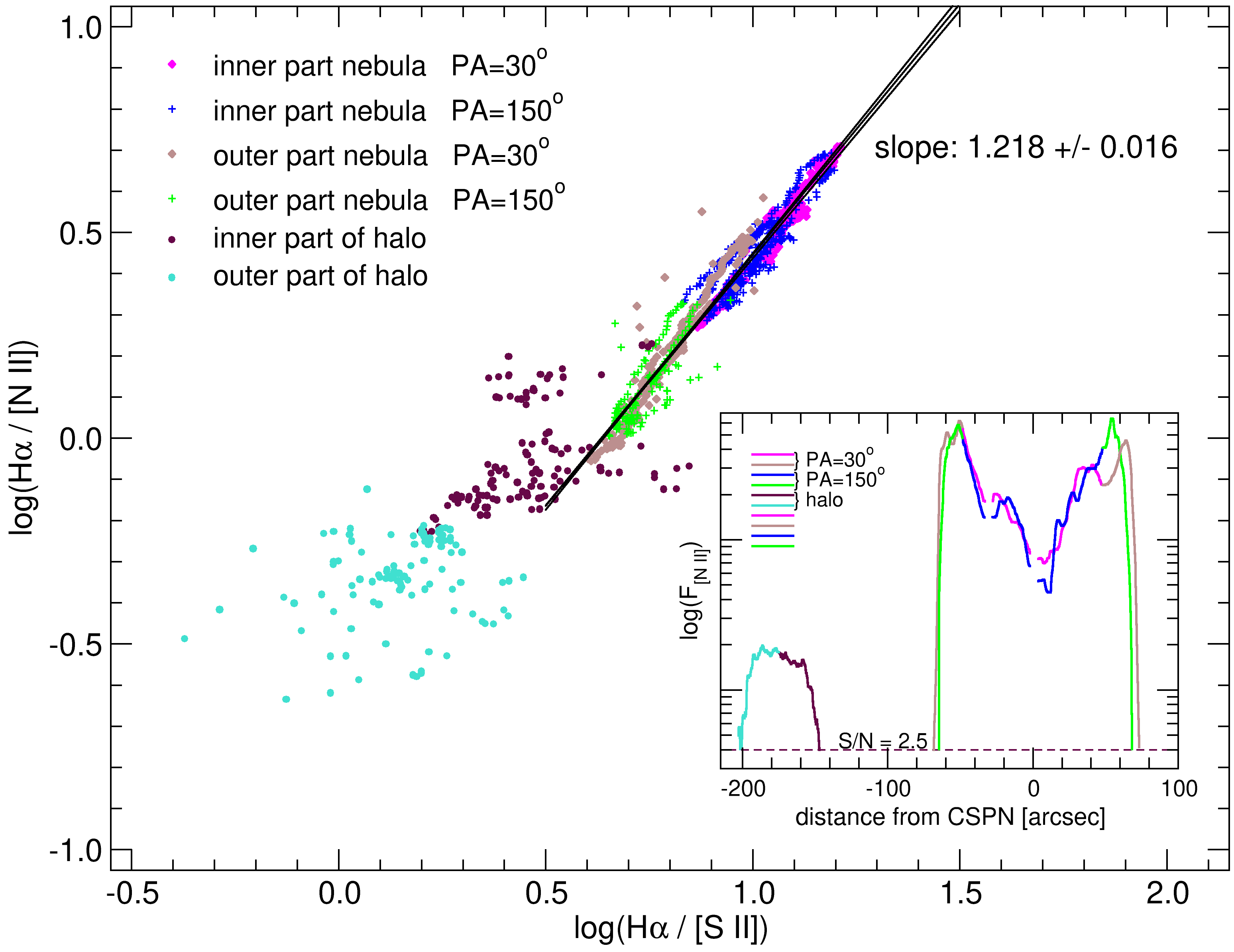}}
\caption{The extended 2D diagnostic diagram of the nebula along the two slits positions and out to the halo in the regions with S/N $>3$ for all lines included in the graph.}
\label{fig_magrini}
\end{figure}

\begin{figure}[t]
\centerline{\includegraphics[width=86mm]{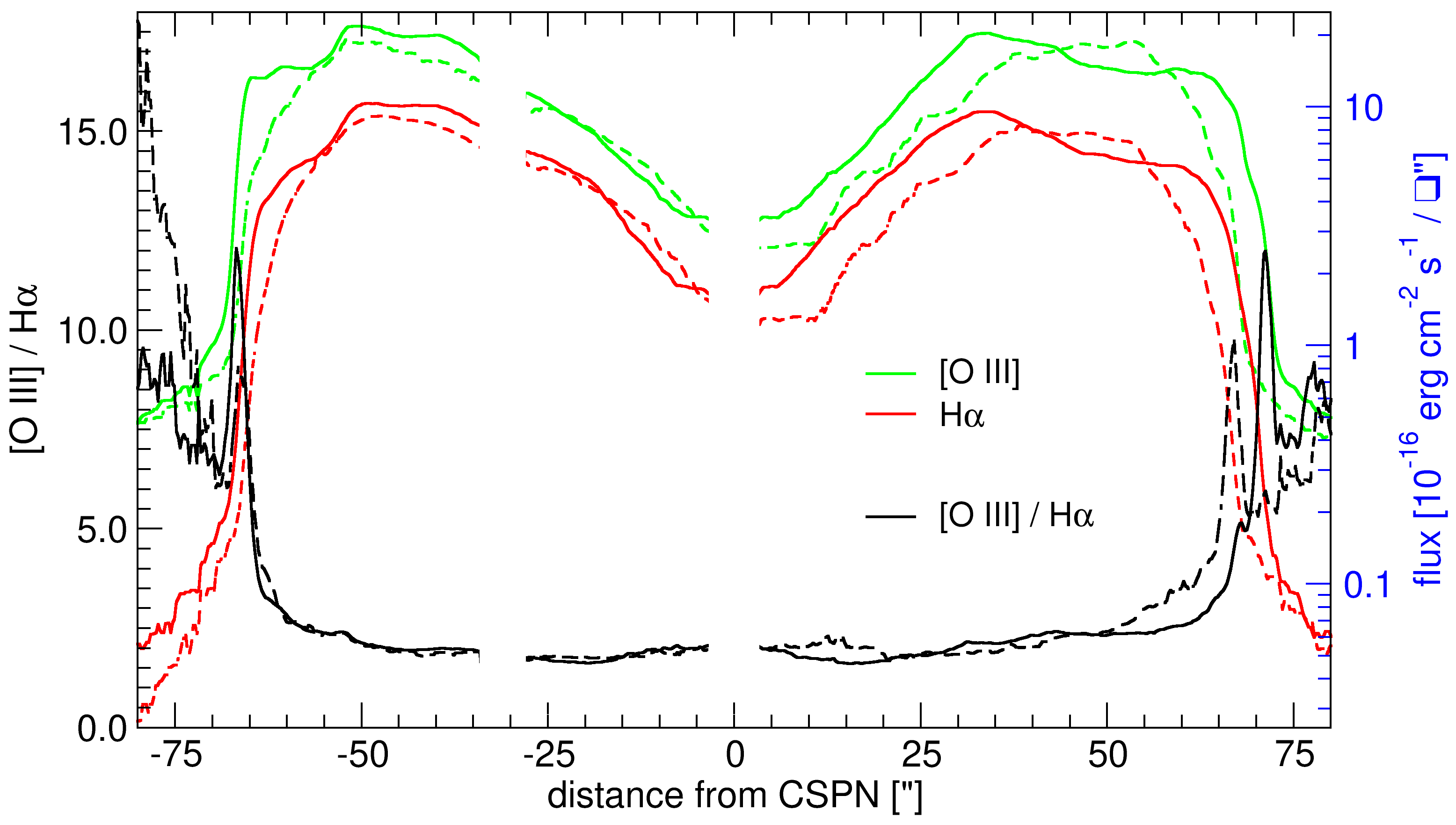}}
\caption{Shock front analysis along the slit in the two FORS2 spectra using $\le$[\ion{O}{iii}]/\ion{H}{$\alpha$}. The spectra were folded with a PSF derived from the central star to obtain same resolution. Solid and dashed  lines indicate for the 30$\degr$ and 150$\degr$ slit spectra, respectively.}
\label{fig_guerrero_shock_spec}
\end{figure}

\begin{figure}[t]
\centerline{\includegraphics[width=86mm]{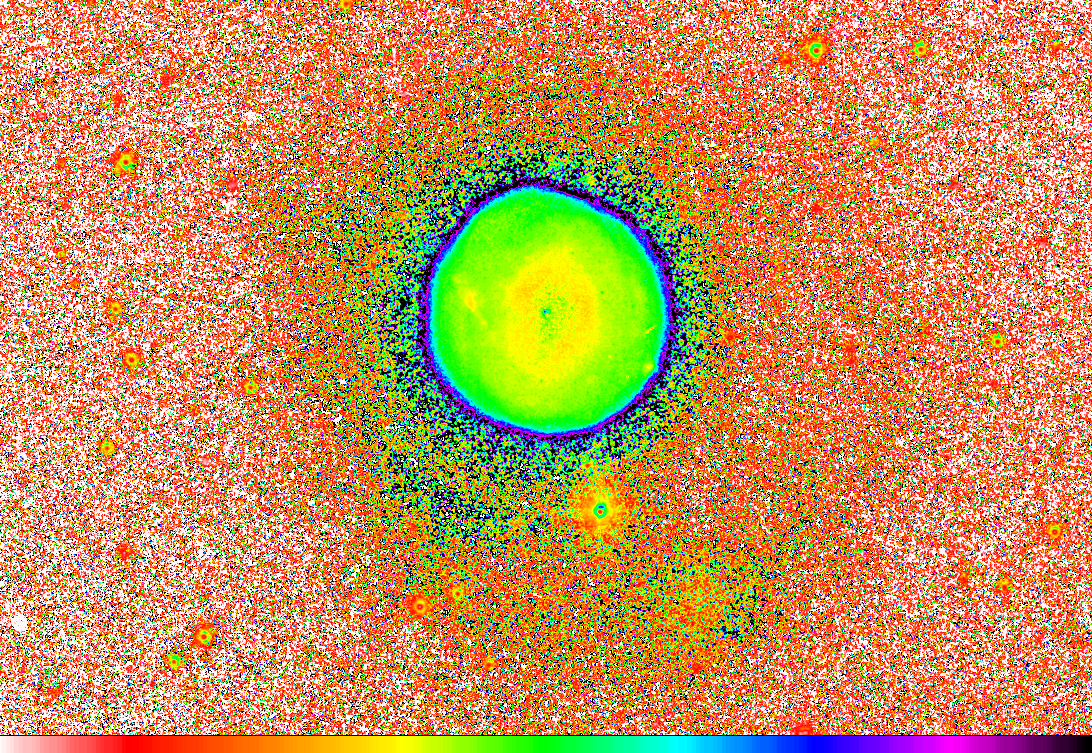}}
\caption{Shock front analysis as in Fig.\ \ref{fig_guerrero_shock_spec} but using the CDK20 narrow band images. The color coding is linear for  0$\le$[\ion{O}{iii}]/\ion{H}{$\alpha$}$\le$5. }
\label{fig_guerrero_shock_image}
\end{figure}

The line ratios $\log$(H$\alpha$/[\ion{N}{ii}]) versus $\log$(H$\alpha$/[\ion{S}{ii}]) have been used in the past to identify photoionized PNe from that excited by additional mechanisms, such as found in supernova remnants or in \ion{H}{ii} regions. First introduced by \citet{garcialario91}, this scheme was refined in \citet{magrini03} and extensively reviewed by \citet{frew10}. \citet{oettl14} extended this analysis to two dimensions to use it for the study of NGC~2438. A more detailed analysis was carried out by us recently using the multiple shell PNe NGC~3242, NGC~6826 and NGC~7662 \citep{Daniela18}. {After investigated on their so called extended 2D diagnostic diagrams (E2DD) we concluded that major deviations from the inside to outside ionization tendency appeared only at regions were FLIERs or LISs are located. Previous studies on excitation properties of FLIERs and LISs \citep[see e.g.][]{akras16, akras17}, support for this scenario.
Regardless no FLIERs or LISs are visible in IC\,5148, we follow the same strategy and studied the nebular ionization tendency through the main fragmentation structures in the nebula by means of its E2DD.} As is shown in Fig.\ \ref{fig_magrini}, where colors denote for different nebular regions, we do not claim for regions at the main nebula showing significant deviations from the general ionization tendency. The regression line gives for a $1.218\pm0.016$ inclination. The fit does not include lines ratios from the halo structure. {At the small PNe sample investigated in \citet{Daniela18} we noted for a possible correlation between the slope of the fit in the E2DD and the CSPNe temperature. However, the previously studied nebula exhibit similar properties like sub-solar abundances, low expansion velocities, high excitation ranges, similar kinetic ages and the presence of FLIERs/LISs. IC\,5148, on the other hand, shows for a higher velocity expansion, no FLIERs/LISs, a solar abundance and highly evolved. Despite the fact that IC\,5148 follow the tendency we found in the previously studied nebulae (higher CSPN/higher slope of the E2DD fit), this result has to be treated with caution, until a further study on the E2DD slope values considering a bigger and diversified sample of PNe could be conducted. }
\\
Following the ideas presented in \citet{Guerrero13} to identify for shock heated regions, we derived the [\ion{O}{iii}]/\ion{H}{$\alpha$} ratios in the FORS2 spectra as well as in the CDK20 images.
As the seeing is a function of wavelength, we used the CSPN to derive a variation of the PSF. Considering the S/N ratio on the central star was very high, we were able to identify not only the central component, but also the wide wings, known to happen due to rare scintillation events of the air. The combined PSF, giving for an error of far below 1\%, consists of a combined Gaussian with a component covering 94\% of the peak intensity (84\% of the integrated flux) and a wing with a 3.5 times wider FWHM with 6\% of the peak intensity (16\% of the total flux). The width varied with wavelength as expected being about 7\% larger at 500\,nm (1\farcs05) than that we found at 660\,nm (0\farcs98). Nevertheless, the wide wing reaches out to 3\farcs0, containing a significant contribution to the total flux. This is of special importance as the line flux drops in the region of interest by more than a order of magnitude. While the main nebula shows a very homogeneous line ratio around 2.5 (see Fig.~\ref{fig_guerrero_shock_spec}), both, the latter spectral analysis as well as the image analysis displayed at Fig.~\ref{fig_guerrero_shock_image}, claim for shock features revealed by a thin shell of enhanced oxygen emission with line ratios significantly higher than 10. {According to \citet{Guerrero13}, this skin of shock-excited material develops when a fast shell expands into a lower density medium. Moreover here also the Mach number changes.} To test the {reliability} of this result, the data were also folded with 2$\times$ and 3$\times$ of the derived PSF. As the feature in the line ratio gets only slightly weaker, respectively, by 20 and 60\% in the ratio, but still giving a line ratio of 5, we can be sure that it is not an artifact of the data analysis.
Moreover, we find a slight steady increase already in the last 8 to 12\arcsec\ inside the front, where the intensity itself resembles still a nearly constantly high flux. We assume that this is caused by the afterglow of the heating in the post-shock region.
The S/N ratio of \ion{H}{$\alpha$} images as well as for the spectra in the halo is not sufficient to derive reasonable numbers for the regions outside the main nebula.

\subsection{Cloudy model}
\begin{figure}[t]
\centerline{\includegraphics[width=88mm]{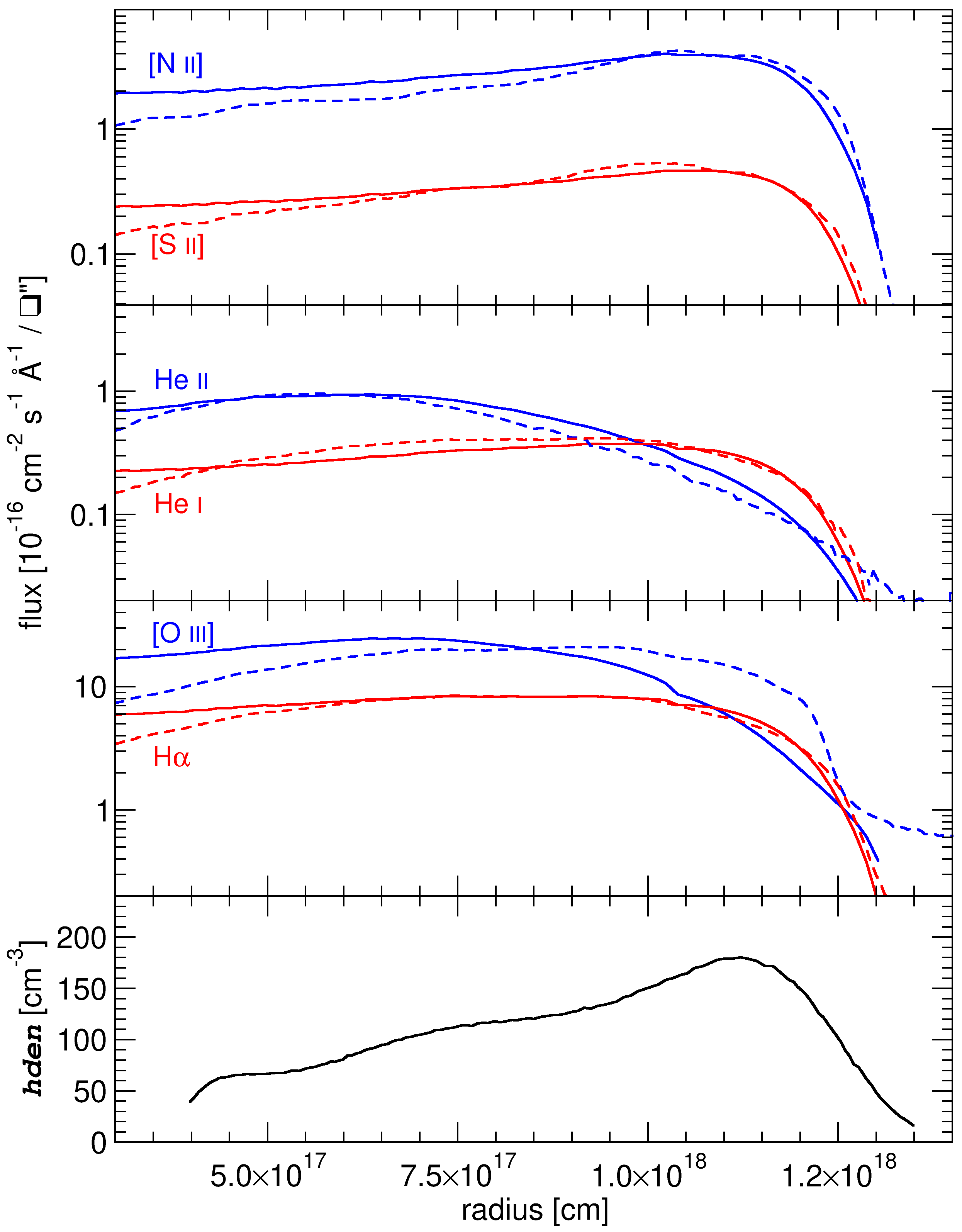}}
\caption{The best fit Cloudy model. Solid lines are giving the model, dashed lines the data averaged in the four directions taken in the slit. In the lowest panel the derived Cloudy {\tt \sl hden} density profile is shown.}
\label{fig_cloudy}
\end{figure}

Physical parameters of the different regions along the main nebula, were investigated by means of Cloudy modeling. The model was performed with version 17.00 of Cloudy \citep{ferland17} and covering only the main nebula. The Cloudy code makes use of an extensive atomic and molecular database together to the possibility to include grains and incorporates a filling factor to simulate clumpiness.\\
The previously derived CSPN parameters such as the effective temperature, luminosity and surface gravity listed in Table\ \ref{tab_cspn}, were used in the model as input parameters for the ionizing source. Thus, we modeled for a central initial mass star of M$_{ini}=1.5\,{\rm M}_\odot$, a metallicity of Z=0.02 dex and a modeled age of 8500 years. The latter measured from the point where the post-AGB object has a $\log(T\rm_{eff})$=3.85 \citep{Miller_Bertolami}. The spectral energy distribution of the CSPN was obtained via state-of-the-art H-Ni NLTE model stellar atmospheres provided by \citet{TMAP}. An expanding spherical geometry was assumed for the nebula. The net line emission provided by Cloudy per volume-unit ($\mathrm{erg\,cm^{-3} s^{-1}}$), was used to create a model 2D projected emissivity profile along the line of sight by an in-house IDL code. In this way, we were able to make a direct comparison of the observed emissivity profiles to the modeled ones. Abundances were set in the Cloudy model to that of a typical metal rich PN as incorporated in the model code from \citet{aller83} and \citet{khromov89}. An assumed GAIA DR2 distance of 1.3\,kpc was used. Although, we also attempt to model the nebula by using the two additional intersection distance solutions found in Sec.~\ref{sec_distance} (1.8\,kpc and 1.5\,kpc). However, the use of these higher distances did not reach convergence for a good model.\\
The modeling was carried out setting an initial tabulated density profile derived from the \ion{H}{$\alpha$} emissivity \citep[see details of this procedure in][]{Daniela18}. As was pointed out in Sec.~\ref{sec_nebular_analysis}, despite the large intensity variation of the [\ion{S}{ii}] lines, the density sensitive line ratio $R_{\rm SII}$ is fairly constant. Hence, we have to assume that the filling factor varies through the nebula. Variations in clumpiness and filling factors were also observed in NGC~2438 in the same way \citep{oettl14} and resolved even in molecular lines in other nearby nebulae \citep[e.g.][and references therein]{matsuura09}.
The estimated line ratio $R_{\rm SII}\sim1.4$ (see Sect.~\ref{sec_nebular_analysis}) give us an upper limit for the electron density of $n_{e}\sim\,200\,\mathrm{[cm^{-3}]}$ \citep{OF06,proxauf14}. It has to be noted here, that the hydrogen density profile in Cloudy is representative for the density within the filled fraction of the nebulosity. In this sense, the integrated nebular mass for such a clumpy model can only be achieved by multiplying the density by the corresponding filling factor in each region.
Based on this initial density profile as a starting point, we manually iterated to convergence for the final model using the observed line profiles of [\ion{N}{ii}]\,658.3\,nm, [\ion{S}{ii}]\,673.0\,nm, \ion{He}{i}\,587.5\,nm, \ion{He}{ii}\,468.5\,nm, [\ion{O}{iii}]\,500.7\,nm and H$\alpha$\,656.3\,nm. The observed profiles were obtained by averaging the four probes taken radially by the two slit directions.
Convergence was derived only by dividing the main nebula into three independent spherical shell regions with independently filling factors as free parameters. The final model was achieved after added the resulting best models for each region (see Fig. \ref{fig_cloudy}).
Final derived Cloudy density profile (called {\tt \sl hden}) is also shown in Fig. \ref{fig_cloudy}.
The resulting parameters for the best model for the nebula are listed in Table \ref{tab_2}.
\\
While we notice very good agreement between the observed and modeled [\ion{N}{ii}], [\ion{S}{ii}], \ion{He}{i}, \ion{He}{ii} and H$\alpha$ emissivity profiles, we observed moderate deviations in the [\ion{O}{iii}] profile. A modeled underestimation on the [\ion{O}{iii}] emission becomes more evident at the outer regions of the main nebula. Considering this, we suggest here for two possible scenarios: one involving emission from a thin hot interclump medium and, a second one related to the residual heating of the post-shock afterglow. The Cloudy code definition of the filling factor assumes that the mass is distributed solely in the clumps. Thus, a possible additional contribution from ionized [\ion{O}{iii}] gas coming from the interclump medium, cannot be considered for our Cloudy clumpy model. Under the second proposed scenario, spectroscopic and image analysis gives evidence for a thin layer of enhanced emission due to a shock front. We might consider then for a possible enhancement of the [\ion{O}{iii}] emission, as a result of the heated post-shock region. This corresponds well also to the previous line analysis shown in Fig.\ \ref{fig_guerrero_shock_spec}.
\begin{table}[t]
\centering
\caption{Main derived nebular properties and nebular parameters obtained from the best Cloudy model. The independent model regions are labeled from in to out as Region I, II and III.}
\label{tab_2}
\begin{tabular}{l l}
\hline\hline
Nebular parameter & Value  \\
\hline
Expansion velocity & 47.9$\pm$1.5\,[km\,s$^{-1}$] \\
Model age & 8\,500\,years \\
Average T$_{e}$ & 11\,500\,[K] \\
Main nebula angular radius &  68$\arcsec$  \\
Main nebula linear radius & $1.32\times10^{18}$ [cm] = 0.43 [pc] \\
Total H mass & 0.34 \,[M$_{\odot}$]\\
& \\
\textit{Region I} & \\
Radius & $(3.98-4.70)\times10^{17}$\,[cm]\\
Filling factor & 0.18 \\
H mass & 0.0015\,M$_{\odot}$\\
\textit{Region II} & \\
Radius &  $(0.47-1.04)\times10^{18}$\,[cm]   \\
Filling factor & 0.37 \\
H mass & 0.15\,M$_{\odot}$\\
\textit{Region III} & \\
Radius &  $(1.04-1.32)\times10^{18}$\,[cm] \\
Filling factor &  0.29 \\
Hydrogen mass & 0.18\,[M$_{\odot}$] \\
\hline
\end{tabular}
\end{table}

\section{Nebular Halo}
A very special set of features in this nebula is occupied by the outer halo. While many nebulae in this evolutionary stage do show one or even more halos, this one has a few very special features to be mentioned in detail.

\subsection{Inner halo}
\begin{figure}[t]
\centerline{\includegraphics[width=88mm]{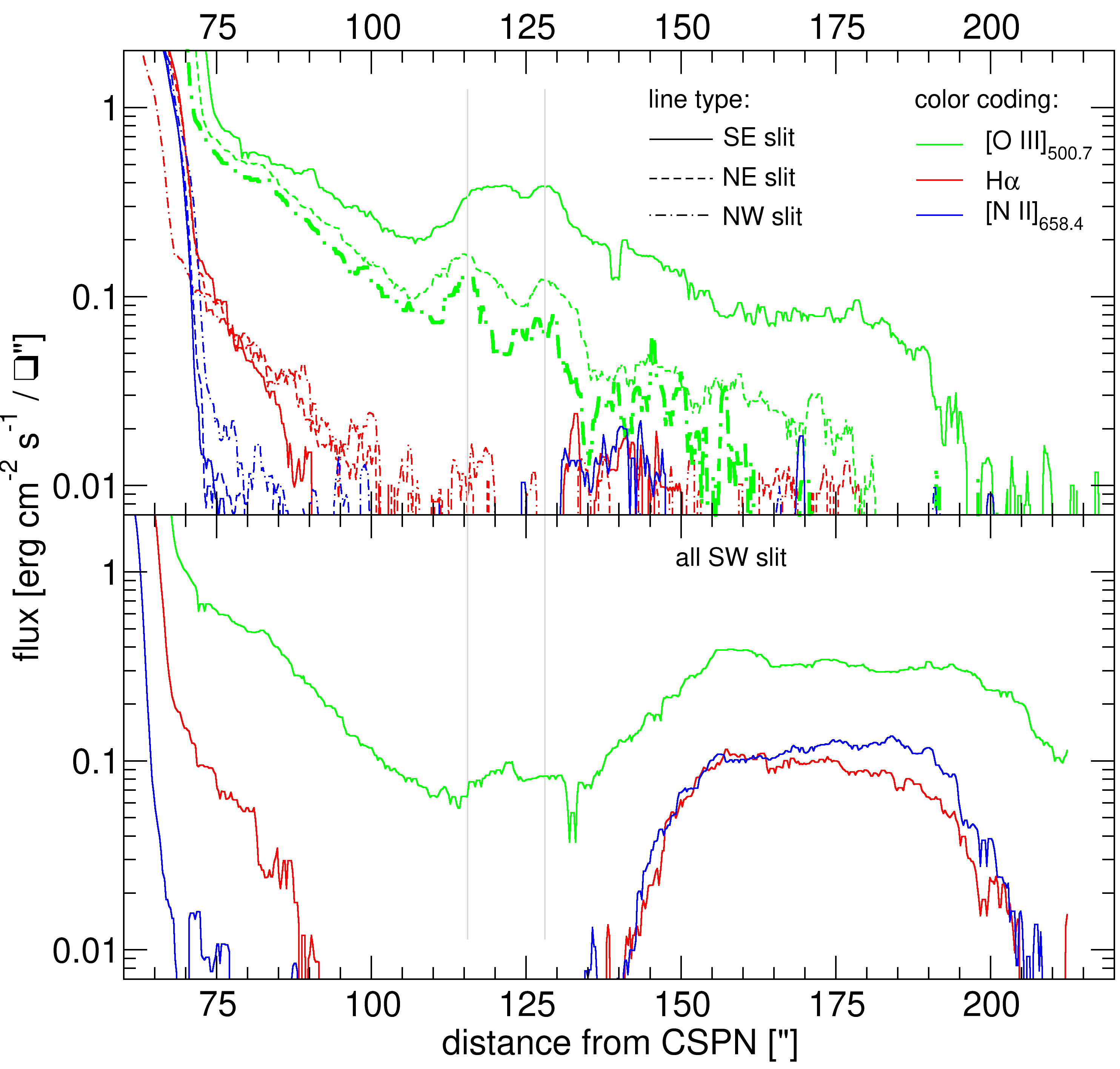}}
\caption{The brightest emission lines in the halo. Upper: the two northern and the southeast section of the halo. Lower: the southwest section. The location of the two halo rings are marked by grey lines.}
\label{fig_halo_lines}
\end{figure}

The inner halo is well represented by an exponential decline of the intensity into nearly all four slit directions equally (Fig.~\ref{fig_halo_lines}). It can be detected in H$\alpha$ and [\ion{N}{ii}](658.4\,nm) up to a radius of about 100\arcsec (corresponding to 0.85\,pc at a distance of 1.3\,kpc) with a constant line ratio of H$\alpha$/[\ion{N}{ii}](658.4\,nm)$\,\approx\,4$. A similar (or only marginally flatter) decline is seen for the [\ion{O}{iii}] line, which is nearly an order of magnitude brighter. Hence, this one can be followed up to the very outskirts (Fig.\ \ref{fig_halo_lines}).
All four slit positions give about the same [\ion{O}{iii}] decline until a radius of 110\arcsec. Beyond that, only the two northern ones, not influenced by the outer halo bow (see Sec. \ref{sec_bow}), seem to continue straight as before and after the halo rings (between 110 and 135\arcsec; see Sec. \ref{sec_ring}). At 175\arcsec (corresponding to 1.5\,pc @$D$=1.3\,kpc) both signals vanishes in the noise. The high [\ion{O}{iii}]/H$\alpha \approx 8$ (even with a slight steady increase towards larger radius) implies that the halo is ionized to a high degree \citep[see modelling and discussion of the halo in NGC\,2438 by][]{oettl14}.
{This ratio and its steady increase is caused by the hardening of the radiation due to the rapid decline of the absorption efficiency of hydrogen }$\kappa_{\rm H} \propto \chi^{-3}$
{for photon energies }$\chi > 1\,{\rm Ryd}$
{ \citep{hardening1}. The effect for the halos of MSPNe is discussed in detail in \citet{hardening2}.  As we cannot measure the density of the halo directly, a
low density $\rho \propto r^{\rm -2}$ was used to test the continuum radiation transmitted by the
CLOUDY model. The incident radiation to the halo is depleted between 1.1 and 2.5\,Ryd to more
than 85\%, while the very hot CSPN having it's radiation maximum above 3\,Ryd is still very strong at
ionization energies of }\ion{He}{ii} {and} \ion{O}{iii}{. The result fit well to this assumption of a photoionized material.
Missing decent density information tracers, the results can used only to follow the tendency and not can be used in absolute quantities.}

\subsection{Halo rings}
\label{sec_ring}
\begin{figure}[ht]
\centerline{\includegraphics[width=88mm]{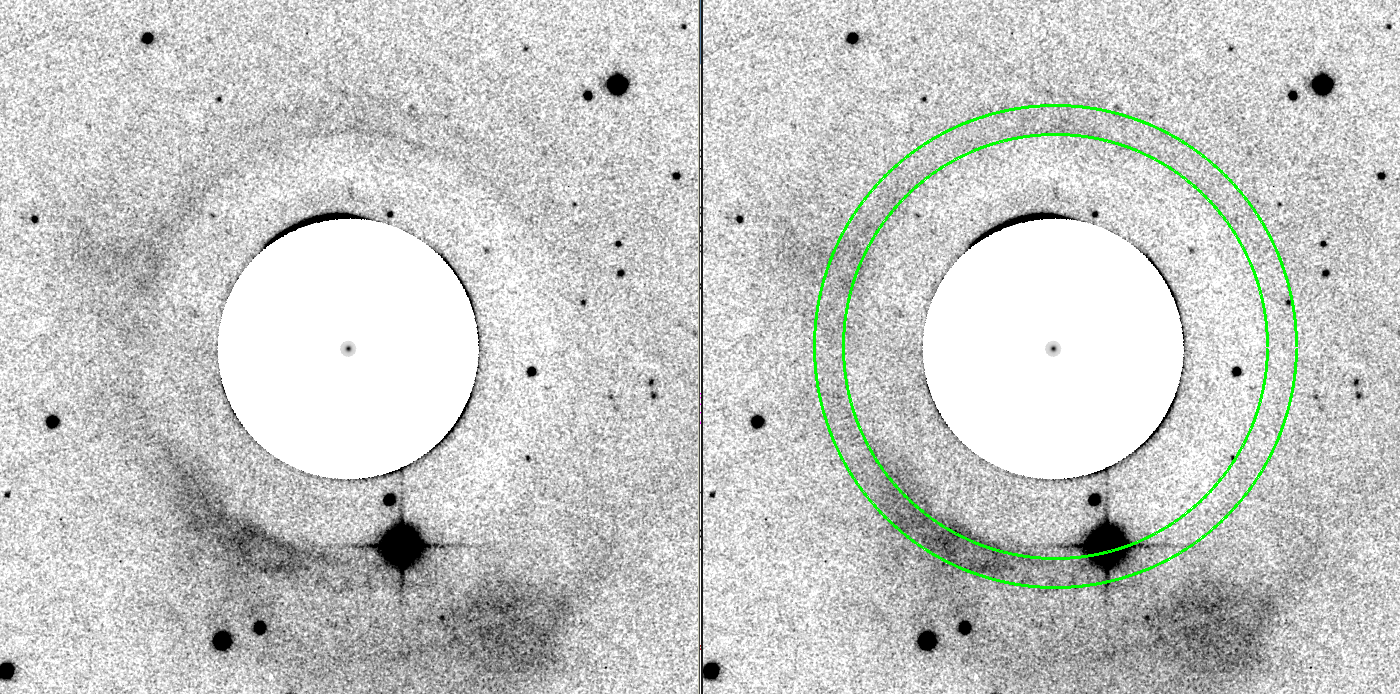}}
\caption{The [\ion{O}{iii}] image of the halo region: Left:  after masking the main nebula and removing a Gaussian shaped decline representing the normal halo intensity distribution. Right: same image with two circles centered exactly at the CSPN at 115 and 128\arcsec.}
\label{fig_ring_image}
\end{figure}
{Multiple outer faint rings and arcs have been found at several planetary nebulae among different morphological types \citep[see e.g.][]{terzian00, corradi04, ramos-larios16}. While several different formation mechanisms have been proposed to explain these structures, their origin is still not that clear, but strongly connected to the mass-loss processes at the latest phase of the AGB. While these features appears scarcely among roundish PNe \citep[$\sim 7$\%,][]{ramos-larios16}, we found in the [\ion{O}{iii}] images of IC\,5148, at a distance of 115 and 128\arcsec, corresponding to 1.00 and 1.08\,pc @$D=$1.3\,kpc respectively, two exact concentric rings centered perfectly respect to the central star. Moreover the images are deep enough to exclude the typically found multiple structures down by nearly an order of magnitude in brighness.} To enhance visibility a power law shaped decline of the regular halo was subtracted from the images (see Fig.\ \ref{fig_ring_image}). The striking is not only the perfect circular shape, but also the fact, that they increase the [\ion{O}{iii}]/H$\alpha$ ratio to at least 30 (lower limit), as these rings are invisible in the other lines. {According to \citet{corradi04}, the spacing between rings correlates with the post-AGB age of the nebula. Although IC\,5148 appears older than the nebulae sample studied by these authors, the large spacing between the rings found here ($\Delta \theta = 13\arcsec, \Delta r= 0.08\,\mathrm{pc} @1.3\mathrm{kpc}$) reinforced for a quite evolved nebula.\\
To date, the most plausible scenario for the formation of such structures is correlated to periodic or quasi-periodic enhancements in the mass-loss rate at the end of the AGB phase. As \citet{ramos-larios16} reviewed, several mechanism(s) have been suggested as responsible for these periodic or quasi-periodic fluctuations such as solar-like magnetic activity in AGB stars \citep{soker00}, long-period oscillations in AGB stars \citep{icke92}, viscous momentum transfer between grains and gas \citep{simis01} or even binary interactions \citep{kim12a,kim12b}. Some others mechanisms as thermal pulses or AGB pulsations have been dismissed basically due to an inconsistency of the time lapses between successive observed rings. But these scenarii produce normally either only a single (or widely spaced) rings or a large number of them. Especially the dust based mechanisms \citep[e.g.][]{simis01} are dominated by reflected light of the main nebula, which would not change the line ratio so much.
Certainly a deep study on the origin of the rings found in IC\,5148 is far beyond the scope of this paper, however as an alternative scenario and following the discussion about shock signatures in \citet{Guerrero13}, we proposed these structures might be caused by shock fronts.} As the other lines, especially also [\ion{S}{ii}] are not detected anymore out there, these features require deeper observations for further verification.

\subsection{Outer halo bow}
\label{sec_bow}
The "bow" is in the outermost part of the halo, peaking into a diffuse clump in the SW region and having a decent visible tail until the SE slit position. It extends from 145 to 200\arcsec\ from the central star, which corresponds to 0.92 to 1.26\,pc at a distance of 1.3\,kpc.
It is at a completely different excitation state than the ring features described above. As shown in Fig.~\ref{fig_halo_lines} it has an H$\alpha$/[\ion{N}{ii}](658.4\,nm) line ratio around unity and [\ion{O}{iii}](500.7\,nm)/H$\alpha$ of about 3. With this respect, it resembles nearly exactly the state of the central nebula and gives an excitation below that found in the innermost halo region (from 70 to 100\arcsec) inside the rings. It continues nicely the excitation direction of the main nebula from inside to outside as shown in Fig.~\ref{fig_magrini}. Thus, it is most likely photoionized material. {Weather it was an unusual asymmetric mass loss event on the AGB, or interaction with interstellar matter (ISM) remains unclear. Although the latter is fairly unlikely at this large distance from the galactic plane. Also the low foreground extinction and the homogeneity of the Balmer line ratio over the whole nebula indicate that a patchy cloudy environment is very unlikely.}

\section{Conclusions}
We performed a detailed photometric and spectroscopic analysis of the high galactic latitude roundish nebula IC~5148, an older evolved nebulae, undisturbed from interaction with the interstellar matter. We found that the distance derived from GAIA DR2 corresponds well to those obtained by statistical methods. Moreover, the very blue color in the GAIA photometry indicates, as shown in \citet{GAIA_PN18}, a good separation of the star from the nebula and give us confidence in the low formal error given by the trigonometric solution of GAIA which results in a distance of $D=1.28^{+0.16}_{-0.13}$\,kpc. Further, we were able to derive a more reliable interstellar foreground extinction leading to E(B-V)=0\fm03 with different independent methods. This is significantly lower than the hitherto often used value of \citet{KSK90} and \citet{KB94}. Medium-high resolution spectroscopy was used to derive a central star temperature of 140$^{+5}_{13}$\,kK. However, after compare to state-of-the-art evolutionary tracks \citep{Miller_Bertolami} we obtain a real value to be slightly lower and thus we adopted 130\,kK and a gravity of $\log(g)\approx7$ on the white dwarf cooling track. The progenitor therefore must has been a low mass star of initially 1.5\,M$_\odot$. This fits well to the expectation that very massive stars should not be found so far from the galactic plane.
We found from our observations of the central star, as well as from the nebular investigation, that IC\ 5148 is, despite its high galactic $z$, non metal underabundant like other nebulae in these regions \citep[e.g. A~15, A~20, MeWe~1-3;][]{Em04}. This was somewhat unexpected and
in contradiction to the postulation of \citet{Soker02}, that spherical PNe might form only in metal poor environment.\\
IC~5148 shows a perfect shock structure at its outermost edge of the main nebula, as predicted in the hydrodynamic models \citep[e.g.][]{Sch04}. It is only visible by using the high ionized species as already suggested and discussed in \citet{Guerrero13}. The shock sensitive diagrams, using low ionized species following \citet{magrini03}, are not sensitive to this effect in the outermost regions. \\
The presented CLOUDY photoionization model shows generally a self-consistent model of the nebula. However, the observed excess of the forbidden [\ion{O}{iii}] line could not be modelled in that way. We speculate that this might be an afterglow of the after-shock region, responsible for driving up the less frequent oxygen atoms even by some small rest of overheated helium. A dedicated time dependent radiative transfer code with source terms of shock heating and afterglows would be required here.\\
Moreover the reported discovery of structured halo emission is enigmatic. The inner halo region resembles the perfect circular shape as that of the main nebula, while further out a bow like structure is seen at one side of the nebula only. While the latter shows line ratios as seen from normal photoionized material, the spherical rings have an extremely large [\ion{O}{iii}]/\ion{H}{$\alpha$} ratio. Further, even deeper spectroscopy than ours obtained at a 8m class telescope, would be required to resolve this puzzle.

\begin{acknowledgements}
We would like to thank the referee for his helpful comments.
The study is based on observations made with
ESO Telescopes at the La Silla Paranal Observatory
under programme ID 098.D-0332. This work has made use of data from the European Space Agency (ESA) mission
{\it Gaia} (\url{https://www.cosmos.esa.int/gaia}), processed by the {\it Gaia}
Data Processing and Analysis Consortium (DPAC,
\url{https://www.cosmos.esa.int/web/gaia/dpac/consortium}).
Furtheron this research has made use of the SIMBAD
database, operated at CDS, Strasbourg, France
\citep{simbad}, the NASA/IPAC Extragalactic
Database (NED) which is operated by the Jet
Propulsion Laboratory, California Institute of Technology,
under contract with the National Aeronautics
and Space Administration and has made use
of “Aladin Sky Atlas” developed at CDS, Strasbourg
Observatory, France \citep{aladin}.
Daniela Barr\'ia and this research were financed by the
FONDO ALMA-Conicyt Programa de Astronom\'ia/PCI 31150001 and
W. Kausch is funded by the Hochschulraumstrukturmittel provided by the Austrian Federal Ministry of Education, Science and Research (BMBWF).
\end{acknowledgements}
\newpage
\bibliographystyle{aa}

\end{document}